\documentclass[twocolumn,showpacs,superscriptaddress]{revtex4-1}
\usepackage{hyperref}
\hypersetup{
  colorlinks=true,        
  linkcolor=blue,         
  citecolor=cyan,         
}

\usepackage{graphicx}
\usepackage{dcolumn}
\usepackage{bm}
\usepackage{color}
\usepackage{enumitem}
\usepackage{amsmath}
\usepackage{amssymb}
\usepackage{orcidlink}
\usepackage{booktabs, tabularx}
\usepackage{makecell}
\usepackage{multirow}
\usepackage{soul}
\setstcolor{red}

\begin{document}
\title{Optical Signatures of a Schwarzschild Black Hole in a Dehnen-Type Dark Matter Halo}

\author{Javokhir Sharipov}
\email{javohirsh100@gmail.com}
\affiliation{Institute of Fundamental and Applied Research, National Research University TIIAME, Kori Niyoziy 39, Tashkent 100000, Uzbekistan}

\author{Jonibek Khasanov}
\email{khasanovjonibek775@gmail.com}
\affiliation{Institute of Fundamental and Applied Research, National Research University TIIAME, Kori Niyoziy 39, Tashkent 100000, Uzbekistan}
\affiliation{National Pedagogical University of Uzbekistan named after Nizami,
Bunyodkor 27, Tashkent, Uzbekistan}

\author{Pankaj Sheoran
}
\email{pankaj.sheoran@vit.ac.in}
\affiliation{School of Advanced Sciences, Vellore Institute of Technology, Tiruvalam Rd, Katpadi, Vellore, Tamil Nadu 632014, India}

\author{Sanjar Shaymatov}
\email{sanjar@astrin.uz}
\affiliation{Institute of Fundamental and Applied Research, National Research University TIIAME, Kori Niyoziy 39, Tashkent 100000, Uzbekistan}
\affiliation{University of Tashkent for Applied Sciences, Str. Gavhar 1, Tashkent 100149, Uzbekistan}
\affiliation{Tashkent State Technical University, 100095 Tashkent, Uzbekistan}

\author{Bobomurat Ahmedov}
\email{ahmedov@astrin.uz}
\affiliation{School of Physics, Harbin Institute of Technology, Harbin 150001, People’s Republic of China}
\affiliation{Institute for Advanced Studies, New Uzbekistan University, Movarounnahr str. 1, Tashkent 100000, Uzbekistan}
\affiliation{Institute of Theoretical Physics, National University of Uzbekistan, Tashkent 100174, Uzbekistan}

\date{\today}

\begin{abstract}
    In this paper, the optical effects that occur near a Schwarzschild-like black hole (BH) with a Dehnen-type $(1,4,2)$ dark matter (DM) halo are explored. We first derive the photon sphere radius and obtain an analytical expression for the deflection angle in the weak-field regime by applying the Gauss–Bonnet theorem (GBT). For the strong-field regime, we perform ray-tracing calculations to examine the behavior of light trajectories and determine the corresponding number of orbits. We further compute the BH shadow and gravitational lensing in a plasma medium and provide constraints arising from the DM halo parameters. We also extend our analysis to weak gravitational lensing within plasma environments, considering both uniform and singular isothermal sphere (SIS) distributions. We find the analytical expressions for the deflection angle in the presence of plasma, and examine the resulting effects on image magnification. The overall results highlight how DM halo properties and plasma characteristics jointly alter observable lensing signatures.
\end{abstract}

\maketitle

\section{Introduction} 

Black holes (BHs) predicted within the framework of general relativity (GR) are fundamentally astronomical compact objects. Modern observations have confirmed their existence, particularly through the shadow images of the supermassive objects M 87* and Sgr A* at the centers of M 87 and the Milky Way galaxies, which were obtained by the Event Horizon Telescope (EHT)~\cite{Akiyama2019b,Akiyama:2022wkp}. These observations have enabled investigations of GR in the strong-gravity regime, while also opening new avenues for studying environments near BHs, including their interactions with dark matter (DM). The BH environment is expected to be formed under the influence of DM. However, there is no interaction of DM with electromagnetic radiation and, therefore,  it can be observed only through gravitational interaction in galaxies and galaxy  clusters 
~\cite{Planck:2018vyg,Rubin1970ApJ}. Developing exact analytical models that describe the influence of DM on the gravitational field of BHs and identifying their observational signatures have become a central direction of contemporary theoretical and observational astrophysics~\cite{Karamazov_2021,Xamidov:2025prl}.

Astrophysical observations strongly indicate that the supermassive  BHs residing in galactic cores are typically enshrouded by distributions of matter, notably including DM halos~\cite{Rees:1984si,Kormendy:1995er,Iocco:2015xga,Bertone:2018krk}. The gravitational influence of DM, while its intrinsic non-gravitational nature remains elusive, is evidenced by its effects on galactic rotation curves, large-scale cosmic structures, and colliding galaxy clusters~\cite{Rubin1970ApJ,Corbelli:1999af,Davis:1985rj}. This motivates the investigation of DM effects in strong-gravity regimes, where their influence may become observationally relevant.  A prevalent modeling approach treats the DM as a spherical halo that envelopes the BH with a symmetric spherically symmetric density profile, serving as a key framework for such systems~\cite{Gondolo1999,Bertone:2005xv,Dehnen1993mnras}. Hence, BHs embedded in DM halos serve as important probes for studying both BH astrophysics and the nature of DM.

DM constitutes roughly 90\% of a galaxy’s mass, with baryonic luminous matter accounting for the remainder~\cite{Persic:1995ru}. This component, which comprises approximately 27\% of the Universe’s content, typically forms extended halos near supermassive BHs~\cite{Planck:2018vyg}. The interaction between DM halos and supermassive BHs is a critical area of research, as DM can significantly influence gravitational dynamics and various astrophysical phenomena. Numerous theoretical models, including the Navarro–Frenk–White~\cite{Navarro:1995iw}, Burkert~\cite{Burkert_1995}, Dehnen-type profiles~\cite{Dehnen1993mnras,Xu:2018wow,Shukirgaliyev2021A&A,Al-Badawi:2024asn,2025EPJC...85.1432U}, and other analytical models  \cite{Cardoso22DM,Maeda:2024tsg,Hou18-dm,Shen24PLB}, have been proposed to describe DM halos, leading to analytical BH solutions that incorporate such environments~\cite{Gohain:2024eer}. In recent years, for exploration of  the influence of DM on optical properties of BHs as BH shadows, deflection angles in gravitational lensing,  and quasinormal  modes of BHs, the Dehnen-type profile for DM has been extensively investigated  
~\cite{Al-Badawi:2025njy,Tahelyani:2024cvk,Uktamov:2025lsq,Li:2025qcv,Al-Badawi25CTP_DM}. The motion of particles and the surrounding spacetime geometry have been profoundly affected by the DM distribution, emphasizing the importance of investigating BHs with the DM halo to advance our understanding of both the nature of DM and strong-field gravity.

Supermassive BHs at the center of galaxies can now be visualized by their shadows—dark shapes set against the glow of surrounding dark matter\cite{1999AAS...195.6207F,2021ApJ...920..155B,2018GReGr..50...42C,2022PhR...947....1P}.  
Recently, the first-ever images of SMBH shadows were captured at the centers of the M87 galaxy and our own Milky Way \cite{Akiyama2019b,Akiyama_2019}.  
Following this historic achievement, BH imaging has rapidly grown into a key field for investigating BHs and telling them apart from other ultra-dense celestial bodies.  
In this article, we also present a description of a BH shadow.

In the context of BHs, gravitational lensing offers a fundamental test of GR and serves as a powerful probe of spacetime geometry near massive compact objects \cite{Bozza:2010xqn}. 
When gravitational lensing is analyzed for the compact objects embedded within DM halos, the deflection angle and magnification of the lensed image exhibit characteristic signatures \cite{Azreg-Ainou:2017obt}.
\cite{Azreg-Ainou:2017obt}. Furthermore, the presence of plasma in realistic astrophysical environments gives rise to frequency-dependent refraction effects \cite{Bisnovatyi-Kogan2010,Atamurotov:2021cgh,2013Ap&SS.346..513M,Bisnovatyi-Kogan:2017kii}, thereby modifying both the brightness of lensed images and the deflection angle.
Recent analytical studies of light propagation in dispersive media have improved our understanding of plasma lensing, providing systematic approaches to investigate frequency-dependent light trajectories and intensity variations in astrophysical plasmas \cite{Perlick:2017fio,Perlick:2023znh,Feleppa:2024vdk}. These developments provide robust methods for analyzing frequency-dependent light trajectories and intensity variations in astrophysical plasmas. Such advances are especially important for interpreting radio observations from facilities such as the EHT, where plasma effects play a significant role. Therefore, considering the plasma effects, whether uniform or non-uniform, is important for enabling reliable comparisons with radio observational data sets such as those from the EHT.

In this work, we examine several optical effects around a Schwarzschild-like BH with a Dehnen-type DM halo with parameters $(\alpha,\beta,\gamma)=(1,4,2)$~\cite{Dehnen1993mnras}. We obtain the photon sphere radius and find an analytical form for deflection angle using the Gauss-Bonnet theorem. In the strong-field regime, ray-tracing method is employed to track photon trajectories. Additionally, we compute the BH shadow radius and analyze the impact of the characteristic density $\rho_s$ and scale radius $r_s$ on observable features. Furthermore, we investigate the effects of plasma environments, modeled as a uniform and singular isothermal sphere, on both the deflection angle and image magnification. 

This paper is outlined as follows: In Section~\ref{Sec:metric}, we conduct an analysis of spacetime geometry and the properties of geodesic motion. Section~\ref{Sec:lensing} is devoted to the theory of gravitational lensing, with subsections dedicated to weak lensing (\ref{subsec:weak}) and strong lensing (\ref{subsec:strong}). In Section~\ref{Sec:shadow}, {we derive the equations of photon motion using the Hamiltonian formalism and analyze the effect of the DM halo on the BH shadow. Section~\ref{sec:plasma} then explores how a plasma medium modifies weak gravitational lensing.} In Section~\ref{Sec:magnification}, we analyze the magnification of images with gravitational lenses. Finally, we summarize our results and present concluding remarks in Section~\ref{Sec:conclusion}.

\section{Spacetime and geodesic analysis}
\label{Sec:metric}

We begin by considering a solution describing a spacetime of a Schwarzschild BH {surrounded by} a Dehnen-type DM halo characterized by the parameters $(\alpha,\beta,\gamma)=(1,4,2)$. The metric describing a static and spherically symmetric BH within a DM halo in Schwarzschild coordinates $(t,r,\theta,\phi)$ is given by \cite{2025EPJC...85.1432U}
\begin{equation} \label{metric1}
    ds^2=-f(r)\,dt^2+\frac{dr^2}{f(r)}+r^2(d\theta^2+\sin^2\theta\,d\phi^2),
\end{equation}
where the metric function takes the form
\begin{equation}
    f(r)=1-\frac{2M}{r}-8\pi\rho_s r_s^2\ln\!\left(1+\frac{r_s}{r}\right). \label{metric}
\end{equation}
Here, $M$, $\rho_s$, and $r_s$ denote the BH mass, the characteristic (or scale) density of the DM halo (fixing the overall normalization of the density profile), and the halo scale radius that sets the transition between the inner $(r\ll r_s)$ and outer $(r\gg r_s)$ regions of the halo, respectively. The third term in $f(r)$ represents the gravitational contribution arising from the Dehnen-type DM halo distribution \cite{Dehnen1993mnras,Mo2010}. From Eq.~(\ref{metric}), the standard Schwarzschild solution is recovered in the limit $\rho_s\rightarrow 0$. The surrounding DM halo is modeled by the generalized Dehnen density profile
\begin{equation}
\rho(r)=\frac{\rho_s}{\left(\dfrac{r}{r_s}\right)^{\gamma}
\left[1+\left(\dfrac{r}{r_s}\right)^{\alpha}\right]^{(\beta-\gamma)/\alpha}},
\end{equation}
where $\gamma$ controls the inner cusp slope, $\beta$ determines the asymptotic fall-off at large radii, and parameter $\alpha$ sets the sharpness of the halo’s inner–outer boundary. The specific choice $(\alpha,\beta,\gamma)=(1,4,2)$ corresponds to a steeply cusped inner density profile $\rho(r)\propto r^{-2}$ and a rapidly decaying outer profile $\rho(r)\propto r^{-4}$, leading to logarithmic corrections in the metric function and representing a dense dark matter environment around the BH.

This spacetime metric enables probing how DM halo impacts BH spacetime geometry. We consider the Lagrangian approach to model the motion of particles  \begin{eqnarray}
    \mathcal{L} &=& \frac{1}{2} g_{\sigma\lambda} \dot{x}^\sigma \dot{x}^\lambda \\&=& \nonumber\frac{1}{2} \left[ - f(r) \dot{t}^2 + f(r)^{-1} \dot{r}^2 +r^2d\dot{\theta}^2+ r^2\sin^2\theta \dot{\phi}^2 \right]\ ,
\end{eqnarray}   
where $\dot{x}^\sigma$=$u^\sigma$=$\frac{d{x}^\sigma}{d\tau}$ ($\sigma$,$\lambda$=0,1,2,3) is the vector of 4-velocity, and $\tau$ is the proper time for massive particles moving along timelike geodesics of timelike and the affine parameter in the case of null geodesics, respectively. Henceforth, motion is constrained to the equatorial plane $(\theta = \pi/2)$ for simplicity. It must be emphasized that the Lagrangian does not depend on coordinates $(t,\phi)$. Hence, the corresponding 4-momenta $(p_{t},p_{\phi})$ are conserved. With this in view, the radial equations of motion for massive and massless particles can be described via the Euler-Lagrange equations
\begin{figure*}[!htb]
 \centering
    \includegraphics[scale=0.55]{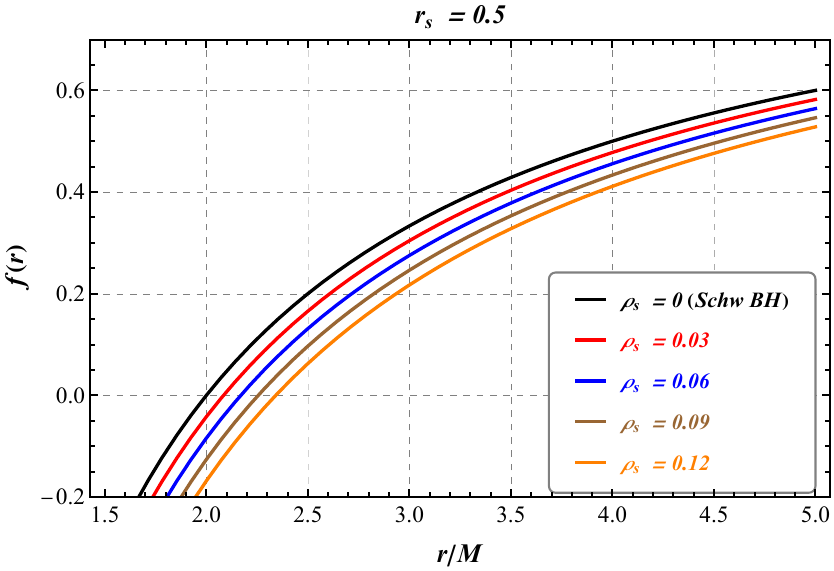}
     \includegraphics[scale=0.55]{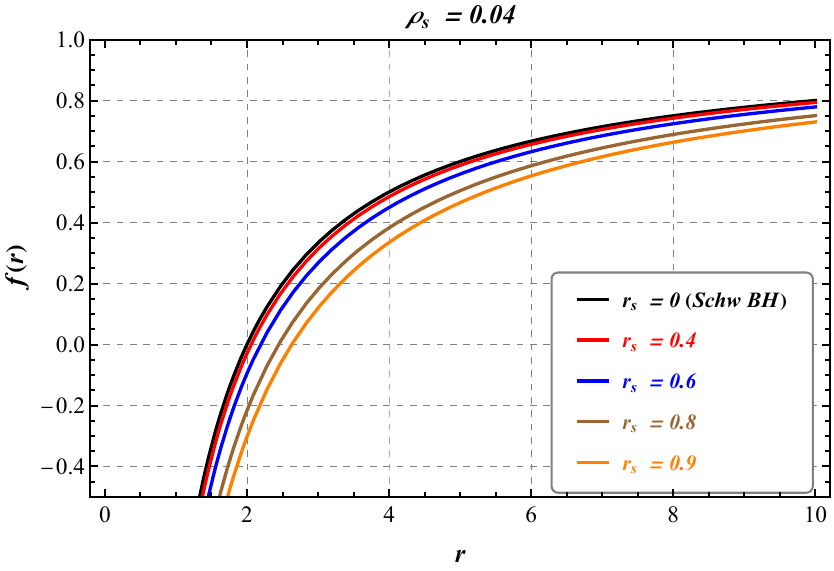}
      \caption{\textcolor{black} The radial profile of $f(r)$ is plotted for various values of $\rho_{s}$ (left) with $r_{s}$=0.5 and $r_{s}$ (right) with $\rho_{s}$= 0.04.}
    \label {fig1}
     \end{figure*}
\begin{equation}\label{euler}
    \frac{d}{dt}\Big(\frac{\partial \mathcal{L}}{\partial  \dot{x}^\sigma} \Big)-\frac{\partial \mathcal{L}}{\partial{x}^\sigma}=0\, ,
\end{equation}
and the normalization condition in terms of the usual relation for the 4-momentum 
\begin{eqnarray}\label{nor}
g_{\sigma\lambda} p^\sigma p^\lambda = k\, .
\end{eqnarray}
It should be noted that for massive particles one has to set $k=-m^2$ with mass $m$ of test particle, while one has to set $k=0$ for photons. Taking Eqs.~(\ref{euler}) and (\ref{nor}) together, one can write the corresponding equations of motion as follows:  
\begin{eqnarray}
    \frac{\partial \mathcal{L}}{\partial \dot{t}}  &=& -f(r) \dot{t}= -E\, ,\\
\label{p}
 \frac{\partial \mathcal{L}}{\partial \dot{\phi}}  &=&L= r^2  \dot{\phi}\, , \\
\label{energyangular}
\dot{r}^2 &=& E^2 - f(r)\left(k + \frac{L^2}{r^2} \right)\, ,
\end{eqnarray}
where $E$ and $L$ respectively refer to constants associated with the total energy and angular momentum of the particle or photon. One can rewrite Eq. (\ref{energyangular}) in terms of the effective potential as follows:
\begin{equation}
 \dot{r}^2+V_{eff}=E^2\, ,
\end{equation}
where the effective potential for massless particles is given by 
\begin{equation}
  V_{eff}=  \left(1-\frac{2M}{r} -8\pi\rho_sr^2_s\log{\left(1+\frac{r_s}{r}\right)}\right)\frac{L^2}{r^2}\, .
\end{equation}

Fig.~\ref{fig1} shows the radial characteristic of the metric function \( f(r) \) for different values of the DM halo parameters \( \rho_s \) and \( r_s \).
As can be seen from Fig.~\ref{fig1}, an increase in \(\rho_s\) and \(r_s\) causes a modification in the form of \( f(r) \) and shift its curves towards larger $r$ compared to the standard Schwarzschild case, resulting in a stronger gravitational effect. According to BH models, the \( f(r) \) profile demonstrates regions within and outside the event horizon, thereby playing an important role in explaining the BH-DM systems. It is obvious that the event horizon increases with increasing DM halo parameters, \(\rho_s\) and \(r_s\).  
\begin{figure*}[htb!]
\centering
\includegraphics[scale=0.5]{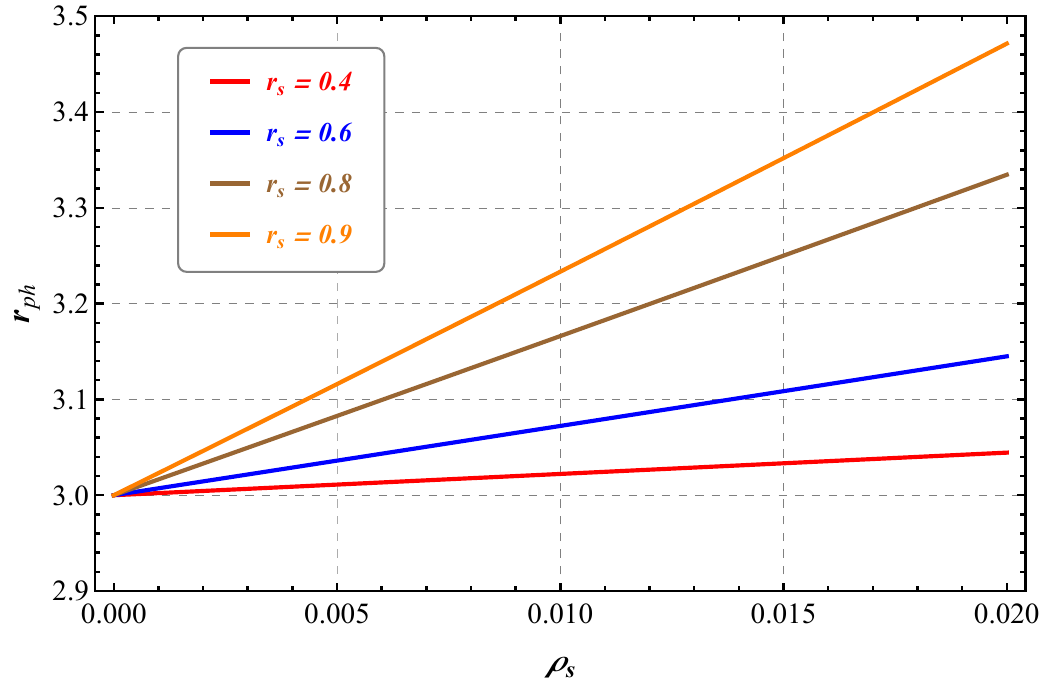}
\includegraphics[scale=0.5]{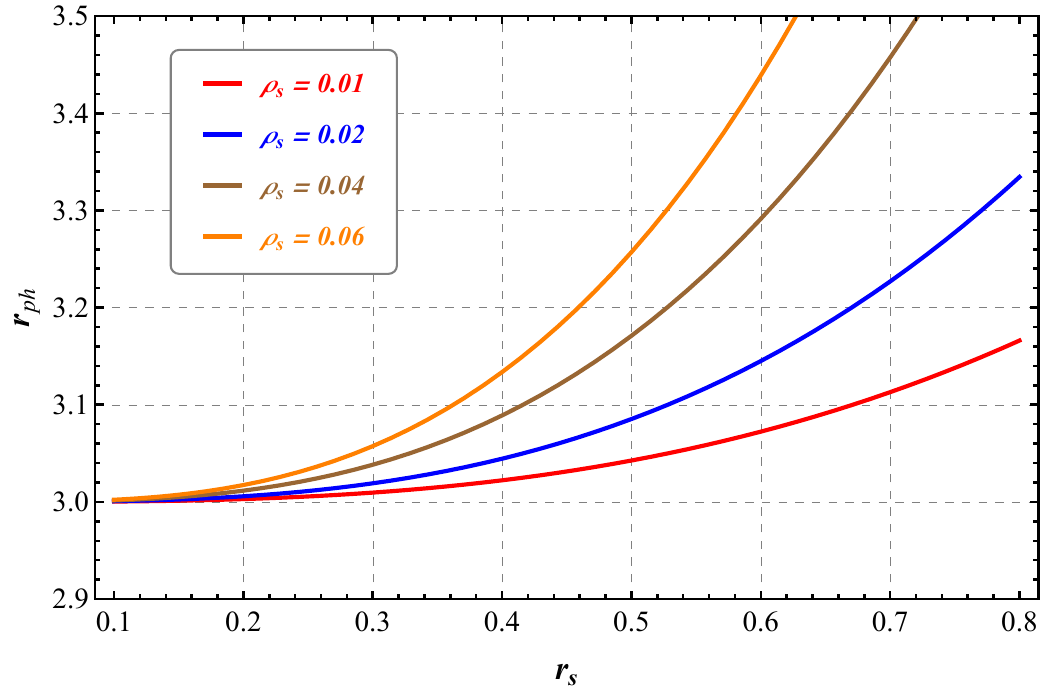}
\caption{Dependence of the photon sphere radius $r_{\text{ph}}$ on the dark matter halo density $\rho_s$ (left panel) and the characteristic scale $r_s$ of the halo (right panel). The figures illustrate how $r_{\text{ph}}$ varies with different values of $\rho_s$ and $r_s$ in the BH model.}
\label{fig2}
\end{figure*}
     
We numerically analyse the photon sphere by solving the following equations simultaneously
\begin{eqnarray}
  V_{eff}=  f(r)\frac{L^2}{r^2}\, \mbox{~and~}\,
\frac{dV_{eff}}{dr}=0\, .
    \label{dVeff}
\end{eqnarray} 

The resulting radii $r_{ph}$ of the photon sphere are shown in Fig.~\ref{fig2}, determining the distance characteristic of stable photon orbits under the gravitational field of the BH with the DM halo. The left panel shows the relationship between the radius of the photon sphere $r_{ph}$ and the density of the DM halo $\rho_{s}$ at different values of the characteristic scale parameter $r_s$. It is obvious that the radius of the photon sphere increases at large values of the characteristic scale parameter. The right panel shows the relationship between the radius of the photon sphere $r_{ph}$ and the characteristic scale factor $r_s$ at different values of the DM halo density parameter $\rho_{s}$. Similarly, we observe that, at large values of the DM halo density parameter, the radius of the photon sphere also increases, as seen in Fig.~\ref{fig2}.

\section{GRAVITATIONAL LENSING}
\label{Sec:lensing}
\subsection{Weak lensing}
\label{subsec:weak}

In this section, we employ the Gauss–Bonnet theorem (GBT) to calculate the light deflection angle in the weak-field approximation for rays propagating near a BH with the DM halo. The GBT establishes a connection between the region $\Omega_{\mathcal{R}}$ and its boundary $\partial\Omega_{\mathcal{R}}$ through the Euler characteristic $\Upsilon(\Omega_{\mathcal{R}})$ that is  the intrinsic geometry of a surface is connected using its curvature 
~\cite{Gibbons2008,Mandal2023,Ovgun2018,Al-Badawi2024,Jusufi:2017}. The expression for this is as follows:

\begin{equation}
\iint_{\Omega_{\mathcal{R}}} \mathcal{K}  dS + \oint_{\partial\Omega_{\mathcal{R}}} \mathfrak{h}  dt + \sum_{z} \alpha_{z} = 2\pi \Upsilon(\Omega_{\mathcal{R}})\ ,
\label{eq:GBT}
\end{equation}
 where $\mathcal{K}$, $dS$, and $\mathfrak{h}$ denote the Gaussian curvature, the area element, and the geodesic curvature, respectively. The sum of the exterior angles is~\cite{Gibbons2008}:
\begin{equation}
 \sum_{z} \alpha_{z} = \alpha_o+\alpha_s\ ,
\end{equation}
where $\alpha_o$ and $\alpha_s$ are the exterior angles at the observer point and the source point, respectively.   

In the equatorial plane ($\theta = \pi/2$), we describe the motion of massless particles with the optical metric, which is derived from the null geodesic condition $ds^2 = 0$. Its line element is given by
\begin{align}
dt^2 &= \tilde{g}_{ij} d\mathbf{x}^i d\mathbf{x}^j = dr_*^2 + \mathcal{F}^2(r_*) d\phi^2\ ,
\label{eq:Metric2}\end{align}
where\\
\begin{align}
dr_*&=\frac{dr}{f(r)}\, , \qquad 
\mathcal{F}(r_*(r)) = \frac{r}{\sqrt{f(r)}}\ \label{eq:NewMFa}.
\end{align}
The determinant $\tilde{g}$ and the nonzero Christoffel symbols for the line element in Eq.~\eqref{eq:Metric2} are

 \begin{align}
 &\tilde{g}=det(\tilde{g}_{ij})=\mathcal{F}^2(r_*)\ ,\\
&\Gamma^{r_*}_{\phi\phi} = - \mathcal{F}(r_*) \frac{d \mathcal{F}(r_*)}{d r_*}\ , \\
&\Gamma^{\phi}_{r_* \phi} =\Gamma^{\phi}_{\phi r_* }= \frac{1}{\mathcal{F}(r_*)} \frac{d \mathcal{F}(r_*)}{d r_*}\ .
\end{align}
By employing these equations, we derive the Gaussian curvature~\cite{Ovgun2018}:
 
\begin{align}
\mathcal{K} =&- \frac{R_{r_*\phi r_*\phi}}{\tilde{g}}= \nonumber- \frac{1}{\mathcal{F}(r_*)} \frac{d^2 \mathcal{F}(r_*)}{d r_*^2} \\\nonumber
 =& - \frac{1}{\mathcal{F}(r^*)} \Big[ \frac{dr}{dr^*} \frac{d}{dr} \left( \frac{dr}{dr^*} \right) \frac{d\mathcal{F}(r^*)}{dr}+ \left( \frac{dr}{dr^*} \right)^2 \\&\times\frac{d^2 \mathcal{F}(r^*)}{dr^2} \Big]\ . \label{eq:gaussC}
\end{align}
Substituting Eqs.~\eqref{metric} and~\eqref{eq:NewMFa} into Eq.~\eqref{eq:gaussC} and expanding up to order $\mathcal{O}(1/r^5)$, we obtain the following result:
\begin{align}
    \mathcal{K}\approx&-\frac{2 M}{r^3}-\frac{8 \pi\rho_s r_s^3}{r^3}+\frac{3 M^2}{r^4}+\frac{24 \pi  M \rho_s r_s^3}{r^4}+\frac{12 \pi  \rho_s r_s^4}{r^4}\nonumber\\&+\frac{48 \pi ^2 \rho_s^2r_s^6}{r^4}\ ,
\end{align}

In Eq.~\eqref{eq:GBT}, the second integral is evaluated in the limit $\mathcal{R}\to\infty$. In this limit, the geodesic curvature and line element take the forms $\mathfrak{h}\to1/\mathcal{R}$ and $dt\to\mathcal{R}d\phi$, respectively (see e.g.,~\cite{Werner2012,Javed2022}) . The angles satisfy $\alpha_o=\alpha_s=\pi/2$, and the Euler characteristic $\Upsilon$ equals one.
\begin{equation}
    \mathfrak{h} dt = \frac{1}{\mathcal{R}} \mathcal{R}  d\phi=d\phi\, ,
\end{equation}

\begin{equation}
\iint\limits_{\Omega_\mathcal{R}} K  dS + \oint\limits_{\partial\Omega_\mathcal{R}} \mathfrak{h}  dt \overset{\mathcal{R} \to \infty}{=} \iint\limits_{S_\infty} K  dS + \int_0^{\pi + \hat{\alpha}} d\varphi.
\end{equation}
\begin{figure*}[!htb]
 \centering
\includegraphics[scale=0.5]{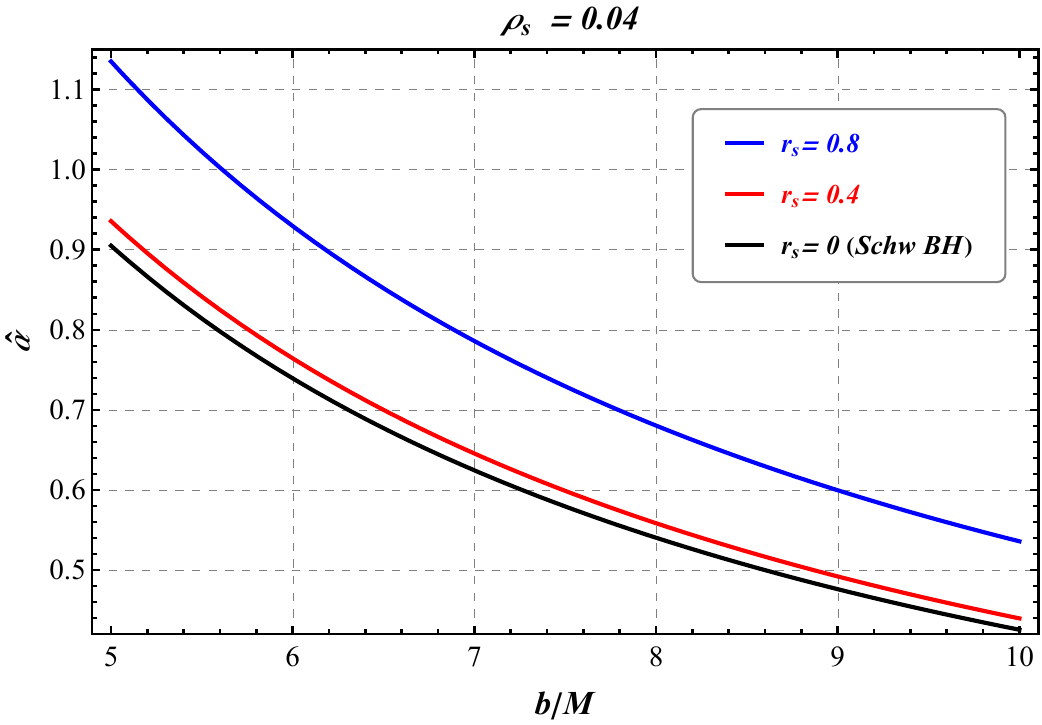}
     \includegraphics[scale=0.5]{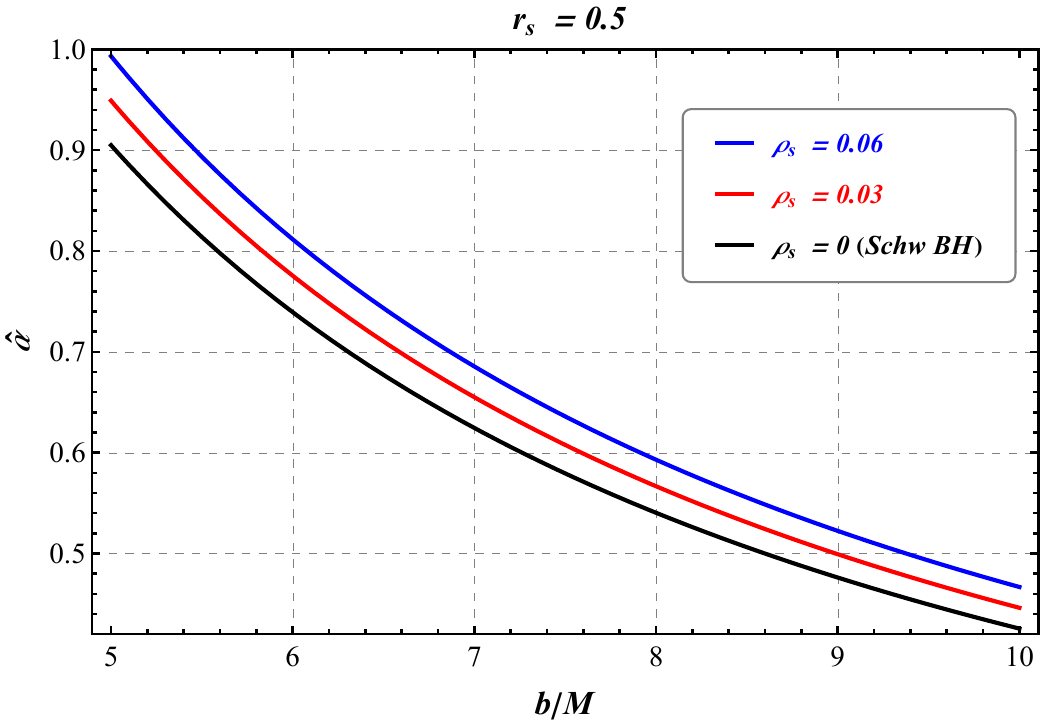}
      \caption{The deflection angle $\hat{\alpha}$ is plotted as a function of the impact parameter $b/M$ for photons.}
    \label{fig:delta}
     \end{figure*}
     
We substitute these into Eq.~\eqref{eq:GBT} and then simplify:
\begin{equation}
\iint\limits_{S_\infty} \mathcal{K}dS + (\pi+\hat{\alpha}) + \pi = 2\pi\ ,
\end{equation}
where $\hat{\alpha}$ is the deflection angle. 
To find the deflection angle, we substitute the area element  $dS = \sqrt{\tilde{g}}\,  dr_*  d\phi$ under the integral. The lower integration limit for $dr$ is set to $r(t)\equiv \mathfrak{B}=\frac{b}{\sin\phi}$ ($b$ is the impact parameter), which represents the polar coordinate expression for a photon's trajectory in a central gravitational field

\begin{align}
\hat{\alpha} =& - \iint\limits_{S_\infty} \mathcal{K}dS = - \int_0^\pi \int_{\mathfrak{B}}^\infty \mathcal{K} \sqrt{\tilde{g}} dr_*  d\phi\nonumber\\ =& - \int_0^\pi \int_{\mathfrak{B}}^\infty \frac{ r \mathcal{K} }{ f(r)^{3/2} }  dr  d\phi\ .
\end{align}

By substituting the Gaussian curvature $\mathcal{K}$ and the metric function $f(r)$ into this equation and simplifying, we obtain the following expression:
\begin{align}
\hat{\alpha} \approx&  \int_0^\pi \int_{\mathfrak{B}}^\infty \Big(\frac{2 M}{r^2}+\frac{8 \pi  \rho_s r_s^3}{r^2}+\frac{3 M^2}{r^3}+\frac{24 \pi  M \rho_s r_s^3}{r^3}\nonumber\\&-\frac{12 \pi  \rho_s r_s^4}{r^3}+\frac{6 M^3}{r^4}+\frac{72 \pi  M^2 \rho_s r_s^3}{r^4}-\frac{48 \pi  M \rho_s r_s^4}{r^4}\nonumber\\&-\frac{45 M^4}{2 r^5}-\frac{360 \pi  M^3 \rho_s r_s^3}{r^5}-\frac{72 \pi  M^2 \rho_s r_s^4}{r^5} \Big)\nonumber\\&\times dr d\phi\ .
\end{align}

 And finally, after integrating, we find the analytical expression for the deflection angle

\begin{align}
\hat{\alpha} \approx&\frac{4 M}{b}+\frac{16 \pi  \rho _s r_s^3}{b}+\frac{3 \pi  M^2}{4 b^2}+\frac{6 \pi ^2 M \rho_s r_s^3}{b^2}-\frac{3 \pi ^2 \rho_ s r_s^4}{b^2}\nonumber\\&+\frac{8 M^3}{3 b^3}+\frac{32 \pi  M^2 \rho_s r_s^3}{b^3}-\frac{64 \pi  M \rho_s r_s^4}{3 b^3}-\frac{135 \pi  M^4}{64 b^4}\nonumber\\&-\frac{135 \pi ^2 M^3 \rho_s r_s^3}{4 b^4}-\frac{27 \pi ^2 M^2 \rho_s r_s^4}{4 b^4}\ .
\label{eq:defA}
\end{align}

The plots of this equation is shown in Fig.~\ref{fig:delta}. In this figure, the black lines correspond to the Schwarzschild BH case. In the left panel, the characteristic density of the DM halo is fixed at $\rho_s = 0.04$, and as the characteristic scale increases, the deflection angle shifts to higher values. In the right panel, the characteristic scale is fixed at $r_s = 0.5$, and as the characteristic density increases, the deflection angle also increases. In general, a larger deflection angle $\hat{\alpha}$ is associated with larger values of the parameters of the DM halo.

\subsection{Strong lensing} \label{subsec:strong}

     \begin{figure*}[!htb]
 \centering
\includegraphics[scale=0.52]{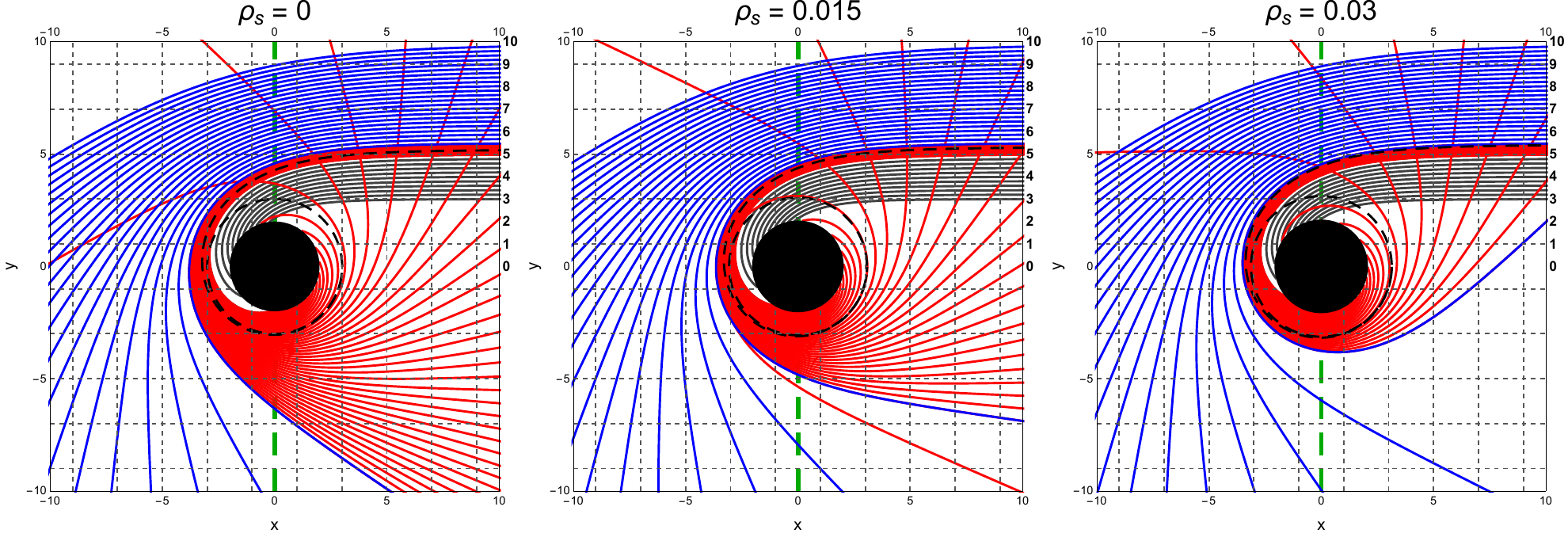}
      \caption{The plots demonstrate ray tracing around a BH with the DM halo, for a fixed value of $r_s=0.5$. The colored curves indicate distinct intervals of the impact parameter $b$. e.g., the gray curve corresponds to $3\le b<5$, the red to $5\le b<5.5$, and the blue to $5.5\le b<10$. The dashed black rings mark the location of the photon sphere. }
    \label{fig:ray}
     \end{figure*}
     \begin{figure*}[!htb]
 \centering
\includegraphics[scale=0.52]{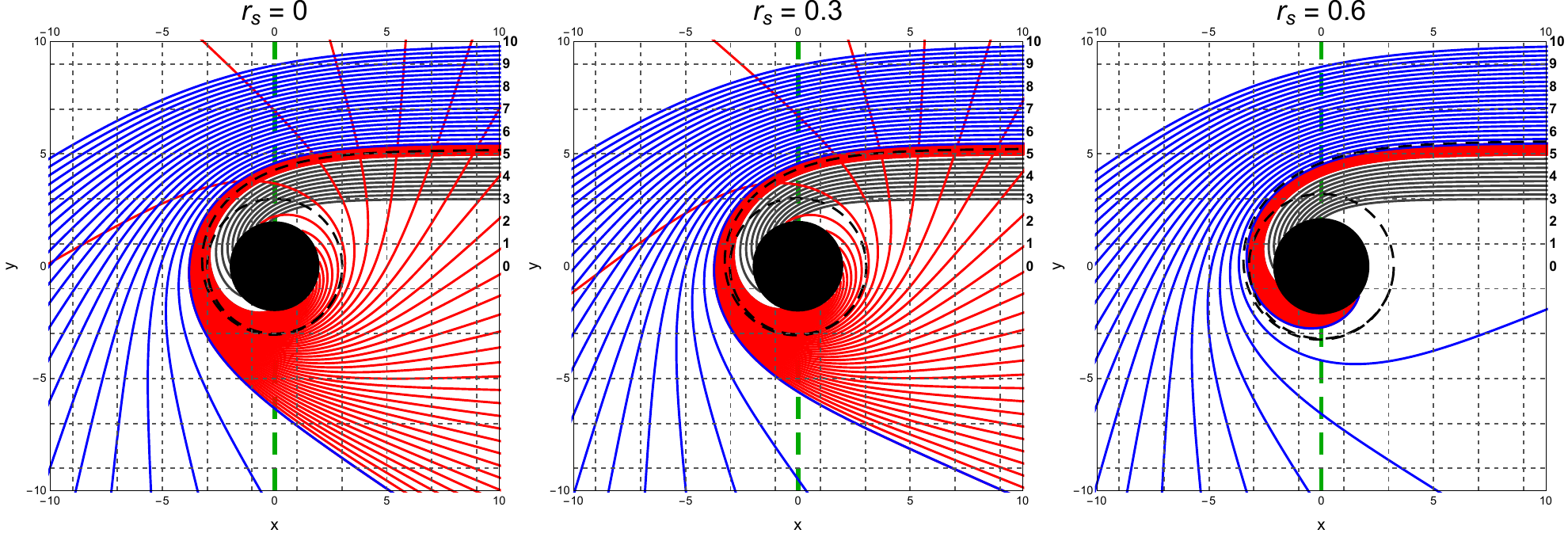}
      \caption{The plots demonstrate ray tracing around a BH with the DM halo, for a fixed value of $\rho_s=0.03$. The colored curves indicate distinct intervals of the impact parameter $b$. e.g., the gray curve corresponds to $3\le b<5$, the red to $5\le b<5.5$, and the blue to $5.5\le b<10$. The dashed black rings mark the location of the photon sphere. }
    \label{fig:ray2}
     \end{figure*}
In this subsection, the photon rings and lensing rings for a BH with the DM halo are analyzed. To support this analysis, we first perform ray-tracing calculations near the BH \cite{Gralla_2019, Cui_2024}.

From Eqs.(\ref{p}) and (\ref{energyangular}), it is possible to derive an orbital equation:
\begin{align}
 \left( \frac{du}{d\phi} \right)^2 =&  \frac{1}{L^2}\left[E^2 - f(u)\left(k+L^2u^2\right)\right] \ ,
\end{align}
\begin{align}
 \left( \frac{du}{d\phi} \right)^2 =&  \frac{E^2}{L^2} - f(u)\left(\frac{k}{L^2}+u^2\right) \ .
 \label{eq:orbit}
\end{align}
Here, we use the transformation $r = 1/u$. By rewriting the orbital equation given in Eq.~\eqref{eq:orbit} for photons ($k=0$), we obtain the following expression:
\begin{equation}
\left( \frac{du}{d\phi} \right)^2 = \frac{1}{b^2} - u^2 f\left(u \right) \equiv G(u)\ ,
\end{equation}
with
\begin{eqnarray}
    G(u)=\frac{1}{b^2}-u^2+2 M u^3+8 \pi  {\rho_s} {r_s}^2 u^2 \log (1+{r_s} u),
\end{eqnarray}
where $b=L/E$ denotes the impact parameter.

The total change in the azimuthal angle $\varphi$ along the trajectory is calculated by integrating the orbit equation \cite{Cai2025}:
\begin{equation}
\varphi = 
\begin{cases} 
\displaystyle
\int\limits_0^{u_h} \frac{du}{\sqrt{G(u)}}\ , \ & if \quad b < b_c\  \\[12pt]
\displaystyle
2 \int\limits_0^{u_{\min}} \frac{du}{\sqrt{G(u)}}\ ,\ &if \quad b > b_c\ ,
\end{cases}
\label{eq:ray}
\end{equation}
where $u_h = 1/r_h$ corresponds to the inverse of the outermost event horizon radius $r_h$, 
$u_{min}$ denotes the smallest positive solution of the equation 
$G(u)=0$.
     
The critical impact parameter is expressed as follows~\cite{hartmann2023gr}:
\begin{equation}
    b_c=\frac{r_{ph}}{\sqrt{f(r_{ph})}}\ ,
\end{equation}
      where $r_{ph}$ is the radius of the photon sphere. 

By solving the integral of the deflection angle $\varphi$ given by Eq.~\eqref{eq:ray}, we obtain the photon trajectories near a BH with the DM halo, as shown in Fig.~\ref{fig:ray}. The characteristic scale is fixed at $r_s = 0.5$, and several values of the characteristic density $\rho_s$ are considered and compared with the Schwarzschild BH case ($\rho_s = 0$).  The black dashed curve corresponds to the trajectory of a light ray at the critical value of the impact parameter, while the dashed ring part represents the photon sphere. 
As the characteristic density increases, the radius of the photon sphere also increases, as can be seen from its crossing of the grid lines. The deflection of the red rays in the figure clearly shows that increasing the characteristic density $\rho_s$ of the DM halo strengthens the gravitational field, leading to larger deflection angles of the photon trajectories.
 Fig.~\ref{fig:ray2} shows results similar to those above; however, in this figure, the characteristic density is fixed at $\rho_s = 0.03$, while the characteristic scale $r_s$ is varied. As the scale $r_s$ increases, the gravitational field becomes stronger. Overall, both parameters of the DM halo contribute to strengthening the gravitational field.

\begin{figure}
    \centering
    \includegraphics[scale=0.6]{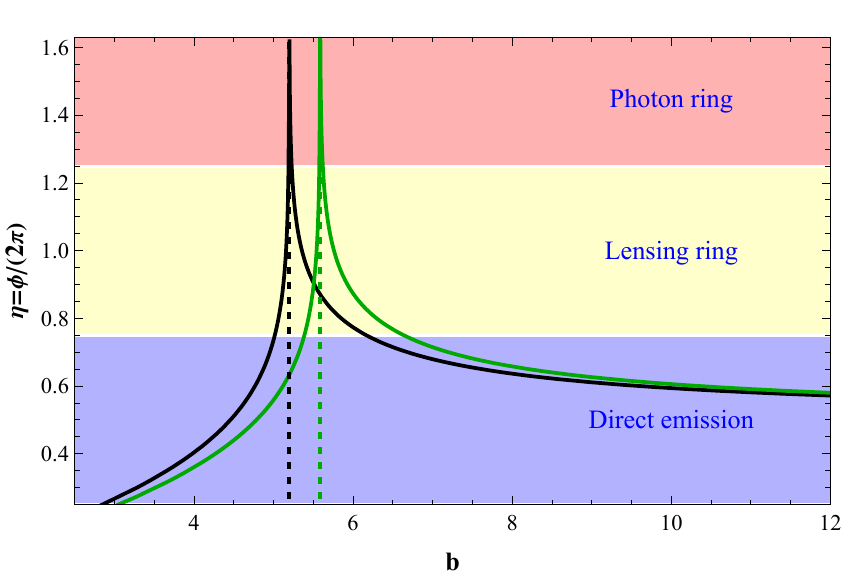}
 \caption{The plot shows the variation of the photon orbit number $\eta$ with impact parameter $b$, where $\varphi$ is the deflection angle. The solid black curve represents a Schwarzschild case, while the solid green curve corresponds to a BH surrounded by a dark‑matter halo with scale radius \( r_s = 0.6 \) and characteristic density \( \rho_s = 0.03 \). Dashed vertical lines mark the critical impact parameters \( b \) for each configuration. The shaded regions correspond to the three observational classes: the photon ring, lensing ring, and direct emission.}
    \label{fig:nb}
\end{figure}

Let us assume that a BH is placed at the origin of the coordinate system. A ray parallel to the $x$-axis in the $xy$-plane comes from infinity and passes near the BH. Owing to the BH's strong gravity, this light ray can intersect the $y$-axis several times. We denote the number of these intersections by $m$. The orbital number $\eta$ is defined as $\eta(b) = \varphi(b)/(2\pi)$. These two quantities are related by:
\begin{equation}
\eta = \frac{2m-1}{4}, \quad m = 1, 2, 3, \cdots\ .
\end{equation}
According to the value of $m$, the rays can be separated into three classes(see e.g.,~\cite{Gralla_2019}): 
\begin{itemize}
  \item Direct emission: $1/4<\eta<3/4$ (in Fig.~\ref{fig:nb}, the blue region)
\item Lensing ring: $3/4<\eta<5/4$ (the yellow region)
\item Photon ring: $\eta>5/4$ (the red region)
\end{itemize}

These classification ranges can also be expressed in terms of the intervals of the impact parameters $b$. The values of these intervals for different characteristic densities $\rho_s$ and scale radius $r_s$ of 
the DM halo are listed in Table~\ref{tab:nb}. From this table, it is clear that as both characteristic parameters increase, all intervals and the critical impact parameter $b_c$ shift towards larger values. For example, for the case ($r_s=0.5$, $\rho_s=0.03$), the lensing ring interval increases to $b\in(5.233,5.414)\cup(5.457,6.443)$, while for the Schwarzschild BH ($\rho_s=0$) it is $b\in(5.015,5.188)\cup(5.228,6.168)$.

Fig.~\ref{fig:nb} shows the number of orbits $\eta$ as a function of the impact parameter $b$. From the comparison of the black and green curves, it can be seen that when a DM halo is present, the peak of the curve shifts toward higher values. Similarly, the critical value of the impact parameter $b_c$ also increases (for the black dashed line, $b_c=5.196$; for the green dashed line, $b_c=5.584$). As the impact parameter $b$ approaches its critical value, $\eta$ tends to infinity. This causes photons to orbit the BH multiple times, forming a photon ring.

\begin{table*}
\centering
\caption{Impact parameter $b$ intervals for different photon trajectory types around BHs with DM halo. $b_c$ denotes the critical impact parameter.}
\label{tab:nb}
\renewcommand{\arraystretch}{1.3}
\begin{tabular}{|c|c|c|c|c|c|}
\hline
\multicolumn{2}{|c|}{\textbf{BHs}} &
\textbf{Direct ($1/4 < \eta < 3/4$)} &
\textbf{Lensing ring ($3/4 < \eta < 5/4$)} &
\textbf{Photon ring ($\eta > 5/4$)} &
\textbf{$b_c$}\\
\Xhline{1.3pt}
\multirow{4}{*}{\rotatebox{90}{\textbf{$r_s=0.5$}}}
& \textbf{$\rho_s=0$} & $b\in(2.848,5.015)\cup(6.168,\infty)$ & $b\in(5.015,5.188)\cup(5.228,6.168)$ & $b\in(5.188,5.228)$\ & $5.196$ \\ \cline{2-6}
& \textbf{$\rho_s=0.01$} & $b\in(2.888,5.088)\cup(6.259,\infty)$ & $b\in(5.088,5.263)\cup(5.304,6.259)$ & $b\in(5.263,5.304)$\ & $5.271$ \\ \cline{2-6}
& \textbf{$\rho_s=0.02$} & $b\in(2.928,5.16)\cup(6.351,\infty)$ & $b\in(5.16,5.339)\cup(5.380,6.351)$ & $b\in(5.339,5.380)$\ & $5.347$ \\ \cline{2-6}
& \textbf{$\rho_s=0.03$} & $b\in(2.969,5.233)\cup(6.443,\infty)$ & $b\in(5.233,5.414)\cup(5.457,6.443)$ & $b\in(5.414,5.457)$\ & $5.423$  \\ \Xhline{1.3pt}

\multirow{4}{*}{\rotatebox{90}{\textbf{$\rho_s=0.03$}}}
& \textbf{$r_s=0$} & $b\in(2.848,5.015)\cup(6.168,\infty)$ & $b\in(5.015,5.188)\cup(5.228,6.168)$ & $b\in(5.188,5.228)$\ & $5.196$ \\ \cline{2-6}
& \textbf{$r_s=0.2$} & $b\in(2.856,5.03)\cup(6.186,\infty)$ & $b\in(5.03,5.203)\cup(5.243,6.186)$ & $b\in(5.203,5.243)$\ & $5.211$ \\ \cline{2-6}
& \textbf{$r_s=0.4$} & $b\in(2.911,5.128)\cup(6.31,\infty)$ & $b\in(5.129,5.305)\cup(5.347,6.31)$ & $b\in(5.305,5.347)$\ & $5.314$ \\ \cline{2-6}
& \textbf{$r_s=0.6$} & $b\in(3.054,5.387)\cup(6.639,\infty)$ & $b\in(5.387,5.575)\cup(5.619,6.639)$ & $b\in(5.575,5.619)$\ & $5.584$  \\ \hline
\end{tabular}
\end{table*}

 \section{BLACK HOLE SHADOW}
\label{Sec:shadow}

 To investigate the trajectories of photons near a BH, we make use of the Hamilton–Jacobi equation. The Hamiltonian describing null geodesics around a BH in the presence of plasma is expressed as follows \cite{Synge1960}
\begin{equation}
\mathcal{H}(x^{\alpha}, p_{\alpha}) = \frac{1}{2} 
\left[ g^{\alpha\beta} p_{\alpha} p_{\beta} 
- \left(n^{2}-1\right) \left(p_{\beta} u^{\beta}\right)^{2} \right]\ .
\end{equation}
Here $x^{\alpha}$ denotes the spacetime coordinates, while $u^{\beta}$ and $p^{\alpha}$ represent the four-velocity and four-momentum of the photon, respectively. It should be emphasized that in the above expression $n$ corresponds to the refractive index, which is defined as $n = \omega / k$, where $k$ is the wave number. It can be written as \cite{Tsupko2009}
\begin{equation}
    n^2=1-\frac{w^2_{p}}{w^2}\ ,
\end{equation}
where $w^2_{p}(x^\alpha)=4\pi e^2N(x^\alpha)/m_{e}$ is the plasma frequency, and  $m_{e}$ and $e$ refer to the electron mass and charge, respectively. N denotes the density of electron numbers.
\begin{figure*}[!htb]
 \centering
\includegraphics[scale=0.36]{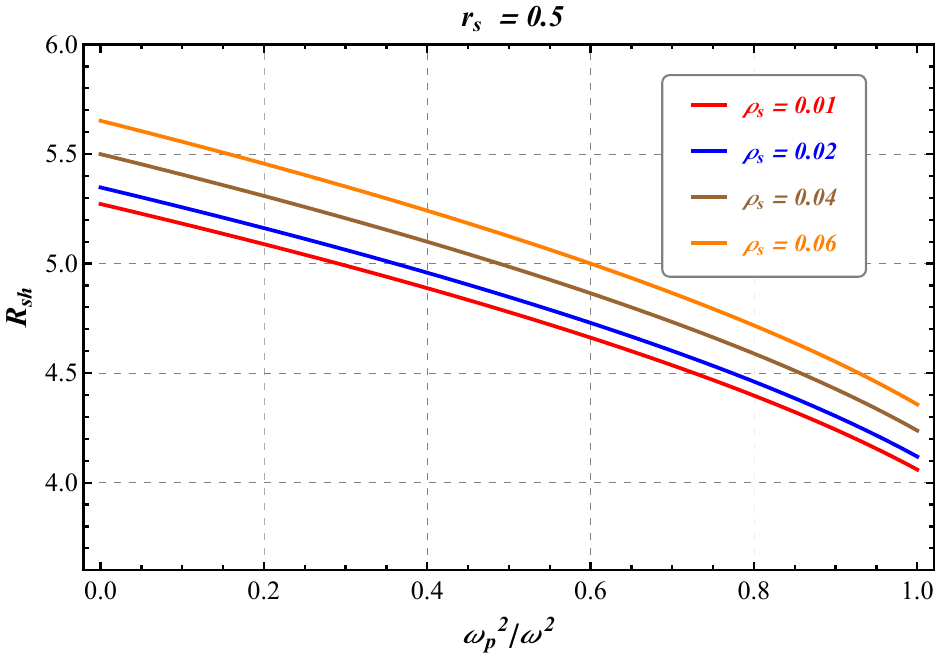}
\includegraphics[scale=0.35]{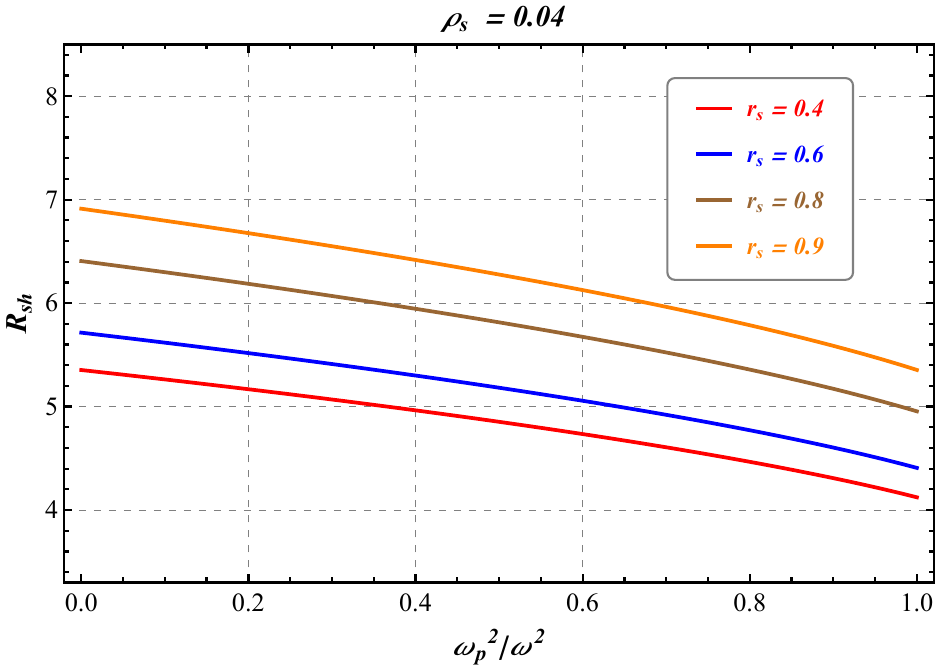}
\includegraphics[scale=0.35]{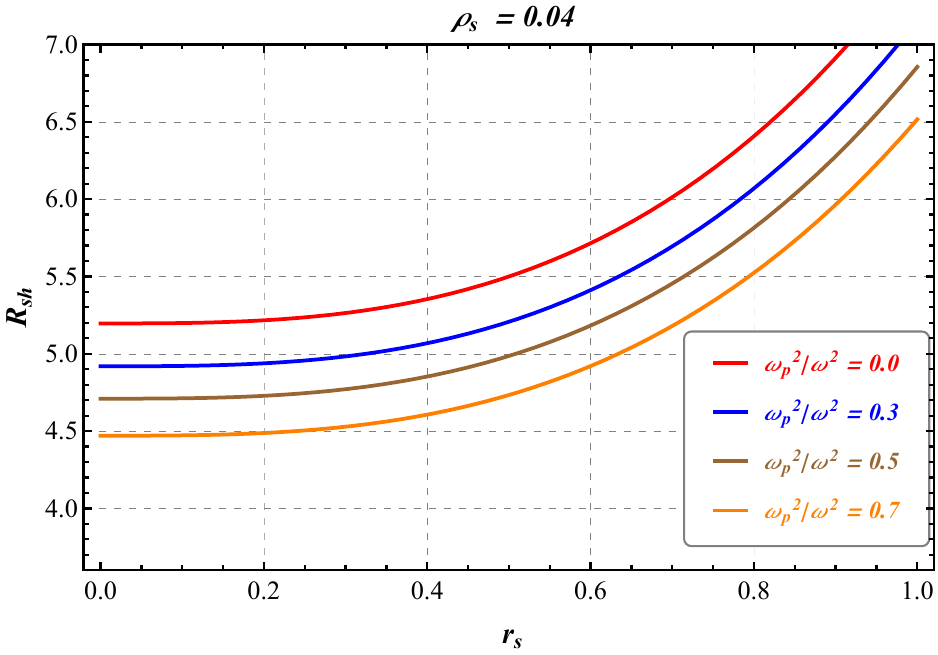}
\caption{Variation of the BH shadow radius with the plasma frequency under different DM halo configurations. 
    (Left panel:) Fixed scale radius \( r_s = 0.5 \) in a different density halo profile. (Middle panel:) For different scale radii, with the DM halo density $\rho_s$ fixed at $0.04$. (Right panel:) Shadow radius as a function of the scale radius \( r_s \) for several plasma frequencies.} \label{fig.7}
\end{figure*}
From the relation $\omega^{2} = (p_{\beta}u^{\beta})^{2}$, the photon frequency can be expressed as
$$
\omega(r) = \frac{\omega_{0}}{\sqrt{f(r)}}, \qquad \omega_{0} = \text{const}.
$$
Here, condition $f(r) \to 1$ holds as $r \to \infty$, ensuring that $\omega(\infty) = \omega_{0} = -p_{t}$ \cite{Perlick_2015}. The Hamiltonian that describes photon geodesics in a plasma medium can then be written as follows \cite{Rogers2015}
\begin{equation}
    \mathcal{H}=\frac{1}{2}\left[ g^{\alpha\beta}p_\alpha p_\beta+w^2_p \right]\ .
\end{equation}
Employing the above relation, the equations that govern light rays in the equatorial plane $(\theta=\pi/2) $ can be expressed as
\begin{equation}
    \dot{t}=\frac{dt}{d\lambda}=-\frac{p_{t}}{f(r)}\ ,
\end{equation}
\begin{equation}
    \dot{r}=\frac{dr}{d\lambda}=p_{r}f(r)\ , \label{eq.33}
\end{equation}
\begin{equation}
    \dot{\phi}=\frac{d\phi}{d\lambda}=\frac{p_{\phi}}{r^2}\ . \label{eq.34}
\end{equation}
Using Eqs.(\ref{eq.33}) and (\ref{eq.34}), one can write
\begin{equation}
    \frac{dr}{d\phi}=\frac{g^{rr}p_r}{g^{\phi \phi} p_{\phi}}\ .
\end{equation}
For light geodesics, where $\mathcal{H}=0$, the above equation can be rewritten as
\begin{equation}
    \frac{dr}{d\phi}=\sqrt{\frac{g^{rr}}{g^{\phi \phi}}} \sqrt{\gamma^2(r)\frac{w^2_0}{p^2_{\phi}}-1}\ ,
\end{equation}
where the following relation is applied: 
\begin{equation}
    \gamma^2(r)=-\frac{g^{tt}}{g^{\phi \phi }}-\frac{w^2_{p}}{g^{\phi \phi}w^2_0}\ . \label{eq.37}
\end{equation}
A photon arriving from infinity reaches its minimum approach at the radius $r_{ph}$, after which it propagates back to infinity. This radius (i.e., the minimum approach) occurs at a stationary point of the function $\gamma^2(r)$. Accordingly, the photon-sphere radius is obtained through the relation below
\begin{figure*}  \includegraphics[scale=0.6]{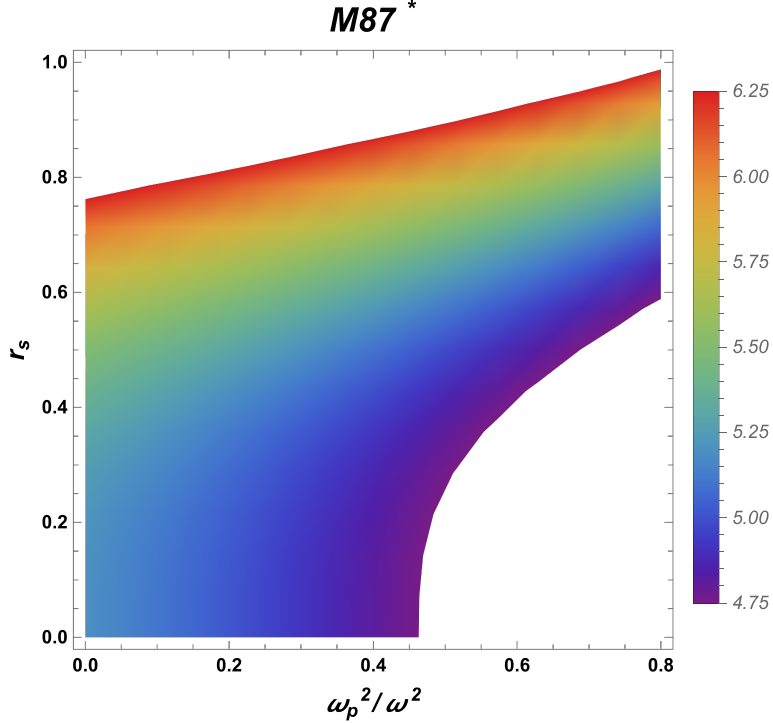}
\includegraphics[scale=0.6]{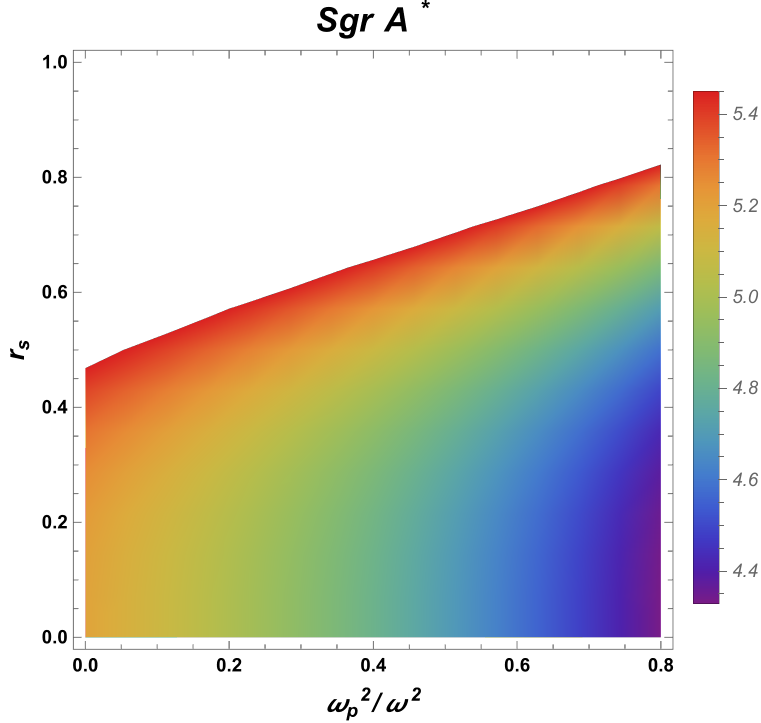}  
\caption {{Allowed values of characteristic scale depending on variation of plasma parameter $\omega_{p}^2$/$\omega^2$ for  SMBHs M 87* and Sgr A*.}}\label{fig.9}
\end{figure*}
\begin{figure*}
\includegraphics[scale=0.6]{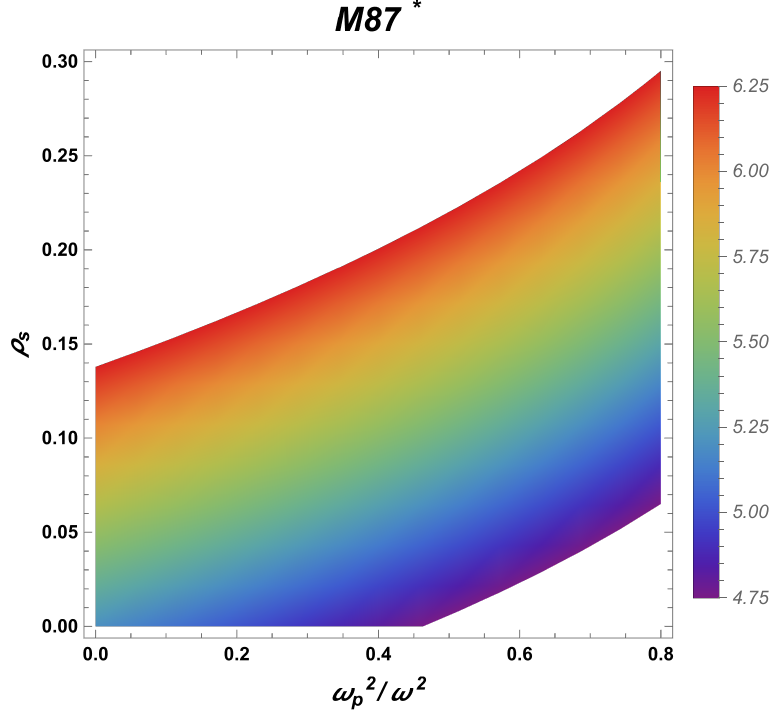}
\includegraphics[scale=0.6]{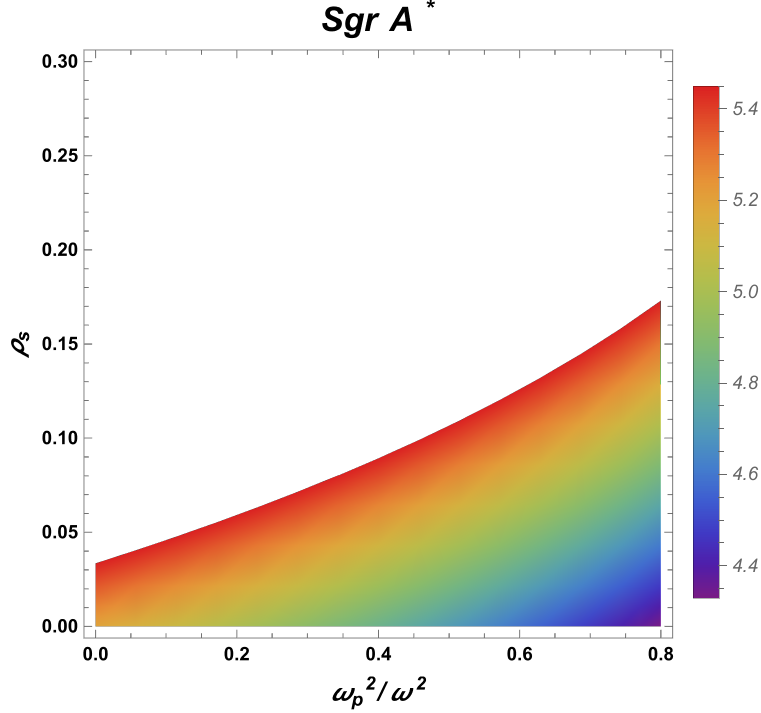}
\caption {Constraints on allowed values of density of the DM halo depending on variation of the plasma parameter $\omega_{p}^2$/$\omega^2$ for the  SMBHs M 87* (left panel) and Sgr A* (right panel).} \label{fig.8}
\end{figure*}

\begin{equation}
\left. \frac{d\big(\gamma^{2}(r)\big)}{dr} \right|_{r = r_{\text{ph}}} = 0 .
\end{equation}
The radius of the photon sphere is examined numerically, and its variation with the DM halo parameters, the scale radius \(r_s\) and the characteristic density \(\rho_s\), is shown in Fig.~\ref{fig2}. As seen from the figure, the radius of the photon-sphere increases under the influence of the DM halo density and the characteristic scale. 

In this section, we study the Shadow of a BH surrounded by DM halo. Firstly, we can write the angular radius $\alpha_{sh}$ of the BH can be obtained as follows\cite{Perlick_2015,KONOPLYA20191}

\begin{align}
\sin^{2} \alpha_{sh} &= \frac{\gamma^{2}(r_{ph})}{\gamma^{2}(r_{0})}= \notag \\[4pt]
&= \frac{
r_{ph}^{2} \left[ \frac{1}{f(r_{ph})} - \frac{\omega_{p}^{2}(r_{ph})}{\omega_{0}^{2}} \right]
}{
r_{0}^{2} \left[ \frac{1}{f(r_{0})} - \frac{\omega_{p}^{2}(r_{0})}{\omega_{0}^{2}} \right]
}\ ,
\end{align}
 where $r_{0}$ and $r_{\mathrm{ph}}$ correspond to the locations of the observer and the photon sphere, respectively. If the observer is located sufficiently far from the BH, the radius of the BH shadow can be approximated using the above equation \cite{KONOPLYA20191}
\begin{align}
R_{sh} &\simeq r_{0} \sin\alpha_{sh}, \notag\\
&= \sqrt{ r^2_{ph} \left[ \frac{1}{f(r_{ph})} - \frac{w^2_{p}(r_{ph})}{w^2_0} \right] }\ .
\end{align}
Based on the asymptotic limit $\gamma(r) \to r$ derived from Eq.~\eqref{eq.37}, the influence of the scalar parameter and the DM halo density on the BH shadow is presented in Fig.~\ref{fig.7}. The figure shows that the shadow radius grows with the characteristic scale $r_s$ of the halo, with the plasma frequency further accelerating this increase. For the compact objects $M87^*$ and $SgrA^*$, we treat them here as static and spherically symmetric, a simplification that differs from the observational constraints reported by the EHT collaboration. Our analysis aims to theoretically constrain the characteristic scale $r_s$ of the DM halo, using observational data provided by the EHT.
\begin{table}[htbp]
\centering
\caption{Observational Data for M87* and Sgr A*. \cite{2024A&A...681A..79E,GRAVITY:2020gka,delaurentis2022accuratemassdistributionm87}}
\resizebox{0.5\textwidth}{!}{
\begin{tabular}{|l|c|c|}
\hline
\textbf{Parameter}         & \textbf{M87*} & \textbf{Sgr A*} \\ \hline
Angular Diameter($\theta$) & $43.3 \pm 2.3 \, \mu\text{as}$ & $51.8 \pm 2.3 \, \mu\text{as}$ \\ \hline
Distance ($D$)             & $16.5  \, \text{Mpc}$ & $8.275   \, \text{kpc}$ \\ \hline
Mass ($M$)                 & $(6.5 \pm 0.7) \times 10^9 \, M_{\odot}$ & $(4.297 \pm 0.013) \times 10^6 \, M_{\odot}$ \\ \hline
\end{tabular}
} 

\label{tab:data}
\end{table}

Based on the constraints for the DM halo scale radius $r_s$ and the plasma frequency, we use observational data from the EHT collaboration—specifically the angular diameter, distance from Earth, and BH mass for both $M87^*$ and $SgrA^*$—listed in Table~\ref{tab:data}. From these values, the shadow diameter per unit mass of the compact object can be computed using the relation~\cite{Bambi_2019}.
\begin{equation}
    d_{sh}=\frac{D\theta}{M}\, .
\end{equation}
Using the relation \( d_{\text{sh}} = 2R_{\text{sh}} \), the BH shadow diameter can be directly expressed. The distance \( D \) is scaled in units of the mass \( M \) \cite{Akiyama2019b,Akiyama2019}, yielding a shadow diameter of \( d^{M87^*}_{\text{sh}} = (11.1 \pm 1.3)M \) for M87* and \( d^{SgrA^*}_{\text{sh}} = (10.1 \pm 0.4)M \) for SgrA*. These results allow us to place bounds on the DM halo parameters: the scale radius \( r_s \) and density \( \rho_s \) as well as on the plasma frequency for the supermassive BHs M87* and SgrA*.  Fig.~\ref{fig.9} displays, via a color map, the allowed regions in the parameter space of the DM halo scale radius \( r_s \) and the normalized plasma frequency \( \omega_p^2 / \omega_0^2 \). The corresponding relation between the DM halo density \( \rho_s \) and the plasma frequency is shown separately in Fig.~\ref{fig.8}.

\section{Gravitational Lensing under Weak-Field Conditions in a Plasma Environment}\label{sec:plasma}

\begin{figure*}[!htb]
 \centering
    \includegraphics[scale=0.5]{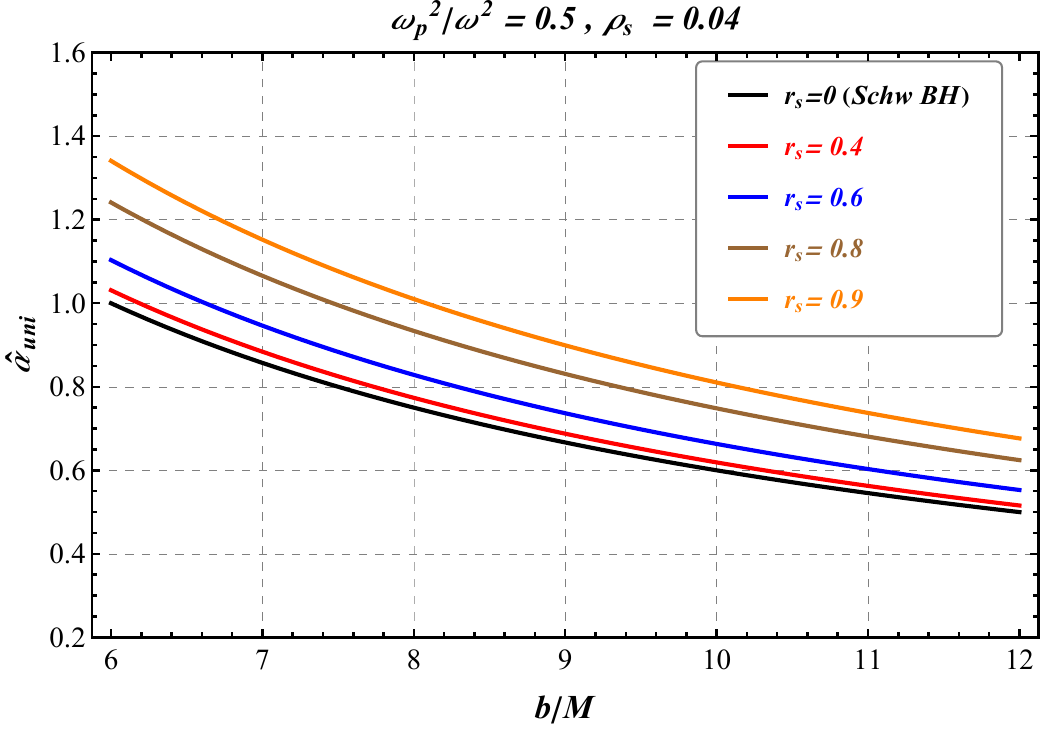}
     \includegraphics[scale=0.5]{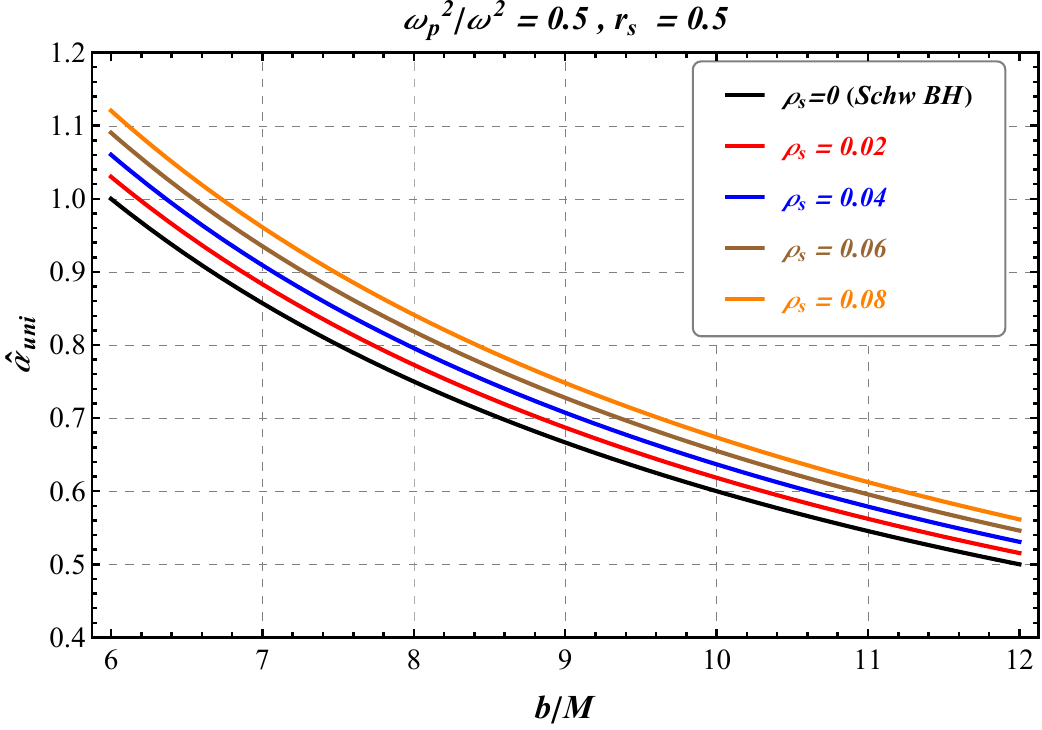}
      \caption{The behaviour of deflection angle $\hat{\alpha }_{\text{uni}}$ versus parameter $b/M$ for different values of scale radius (left panel) and density of DM halo (right panel).}\label{a1}

     \end{figure*}
 This study investigates the weak lensing effects in the vicinity of a DM halo around BH. To that, it is necessary to employ the relevant expression that describes the weak-field approximation (see, e.g.,~\cite{Bisnovatyi-Kogan2010,2021PDU....3200798B}).
\begin{equation}
    g_{\alpha \beta}=\eta_{\alpha \beta}+h_{\alpha \beta}\, ,
\end{equation}
where $\eta_{\alpha\beta}$ denotes the gravitational potential of Minkowski spacetime and $h_{\alpha\beta}$ represents the gravitational field. With this in mind and according to~\cite{Bisnovatyi-Kogan2010}, the relationship between these two potentials is expressed as follows:
\begin{eqnarray}
 &&   \eta_{\alpha \beta}=diag(-1,1,1,1)\ , \nonumber\\
 &&   h_{\alpha \beta} \ll 1, \hspace{0.5cm} h_{\alpha \beta} \rightarrow 0 \hspace{0.5cm} under\hspace{0.2cm}  x^{\alpha}\rightarrow \infty \ ,\nonumber\\
 &&     g^{\alpha \beta}=\eta^{\alpha \beta}-h^{\alpha \beta}, \hspace{0,5cm} h^{\alpha \beta}=h_{\alpha \beta}\, .
\end{eqnarray}
By considering this relation, the deflection angle of light traveling near a BH in the presence of DM halo can be defined as
\begin{eqnarray}
    \hat{\alpha }_{\text{b}}&=&\frac{1}{2}\int_{-\infty}^{\infty}\frac{b}{r}\left(\frac{dh_{33}}{dr}+\frac{1}{1-\omega^2_p/ \omega^2}\frac{dh_{00}}{dr}\right.\nonumber\\&-&\left.\frac{K_e}{\omega^2-\omega^2_p}\frac{dN}{dr} \right)dz\, , 
\end{eqnarray}
where $\omega_{p}$ and $\omega$ represent the plasma and photon frequencies.Under the weak‑field approximation, a Taylor‑series expansion of the spacetime metric leads to the following expression
\begin{eqnarray}
  ds^2=ds^2_0 &+&\left(\frac{2 M}{r}+8 \pi  {\rho _s} {r_s}^2 \log \left(1+\frac{{r_s}}{r}\right)\right)dt^2 \nonumber\\&+&\left({\frac{2 M}{r}+\frac{8 \pi  {\rho_s} {r_s}^3}{r}}{}\right)\frac {z^2}{r^2} dr^2\, , 
\end{eqnarray}
where $ds^2_0=dt^2+dr^2+r^2(d\theta^2+\sin^2\theta d\phi^2)$. The corresponding components of the gravitational perturbation $h_{\alpha \beta}$ in Cartesian coordinates take the form:
\begin{eqnarray}
 h_{00}&=&\left(\frac{2 M}{r}+8 \pi  {\rho _s} {r_s}^2 \log(1+\frac{r_{s}}{r})\right)\, ,\\
 h_{ik}&=&(\frac{2 M}{r}+8 \pi  {\rho _s} {r_s}^2 \log(1+\frac{r_{s}}{r})) n_i n_k\, ,\\
h_{33}&=&(\frac{2 M}{r}+8 \pi  {\rho _s} {r_s}^2 \log(1+\frac{r_{s}}{r})) \cos^2\chi 
\label{h}\, ,
\end{eqnarray}
where  $\cos^2\chi=z^2/(b^2+z^2)$ and $r^2=b^2+z^2$
From these expressions, the first radial derivatives of $h_{00}$ and $h_{33}$  can be written as:
\begin{eqnarray} 
&& \frac{dh_{00}}{dr}=-\frac{2 M}{r^2}-\frac{8 \pi  {\rho_s} {r_s}^3}{r^2 \left(\frac{{r_s}}{r}+1\right)}  ,\\
&& \frac{dh_{33}}{dr}=-\frac{6 z^2 \left(M+4 \pi  {\rho_s} {r_s}^3\right)}{r^4} \ .
\end{eqnarray} 
With these expressions established, we include a plasma medium and investigate the resulting weak gravitational lensing.

\textit{Uniform Plasma:} We begin by defining the deflection angle of light propagating around the BH in a uniform plasma medium, which can generally be expressed as follows~\cite{Khasanov2025JCAP,Alloqulov12025,Alloqulov2024}:
\begin{equation} \label{anglemain}
\hat{\alpha}_{uni}=\hat{\alpha}_{uni1}+\hat{\alpha}_{uni2}+\hat{\alpha}_{uni3}\ ,
\end{equation}
with
\begin{equation} \label{alphaexpansion}
\left.
\begin{aligned}
   \hat{\alpha}_{1} &= \frac{1}{2}\int_{-\infty}^{\infty} \frac{b}{r}\frac{dh_{33}}{dr} \, dz,\\ 
   \hat{\alpha}_{2} &= \frac{1}{2}\int_{-\infty}^{\infty} \frac{b}{r}\frac{1}{1-\omega^2_{e}/\omega^2}\frac{dh_{00}}{dr} \, dz,\\ 
   \hat{\alpha}_{3} &= \frac{1}{2}\int_{-\infty}^{\infty} \frac{b}{r}\Bigg(-\frac{K_{e}}{\omega^2-\omega^2_{e}}\frac{dN}{dr}\Bigg) \, dz.
\end{aligned}
\right\}
\end{equation}
It is worth to note here that one can easily compute the deflection angle numerically from the above Eqs.~(\ref{anglemain}) and (\ref{alphaexpansion}).

Using the obtained results, the dependence of the deflection angle $\hat{\alpha}_{\text{uni}}$ on the $b$ is plotted in Fig.~\ref{a1} for different values of the DM halo parameters $r_{s}$ and $\rho_{s}$, as well as for varying plasma frequencies $\omega^{2}_{e}/\omega^{2}$. The figure illustrates how the deflection angle changes with the plasma parameters for different choices of $r_{s}$ and $\rho_{s}$ (from left to right).  Interestingly, the deflection angle $\hat{\alpha}_{uni}$ is highly sensitive to these parameters, showing a rapid decrease as $r_{s}$ and $\rho_{s}$ vary, as demonstrated in Fig.~\ref{a1}.

\textit{Non-uniform plasma case}: The singular isothermal sphere (SIS) model provides a suitable framework for investigating weak gravitational lensing of photons near a BH in the presence of a non‑uniform plasma. In this model, the spherical gas cloud contains a central point of infinte density described by the density profile \cite{2021PDU....3200798B,Bisnovatyi-Kogan2010}.
\begin{equation}
\rho(r)=\frac{\sigma^2_{\nu}}{2\pi r^2}\ ,
\end{equation}
where $\sigma^2_{\nu}$ is the one-dimensional velocity dispersion, and the plasma concentration follows the analytical form below~\cite{Bisnovatyi-Kogan2010,Bisnovatyi-Kogan:2017kii}
\begin{equation}
N(r) = \frac{\rho(r)}{\kappa \; m_p}.
\end{equation}
Here, $m_p$ and $\kappa$ are the proton mass and a dimensionless constant (i.e. generally associated with the DM halo), respectively. This dimensionless parameter $\kappa$ is frequently connected to the plasma frequency \cite{Bisnovatyi-Kogan2010}.
\begin{equation}
\omega^2_c=K_e N(r)=\frac{K_e \sigma^2_{\nu}}{2\pi \kappa m_p r^2}\ .
\end{equation}
To analyze the effects of non-uniform plasma (SIS), it is necessary to express the deflection angle around the BH. It can be written as~\cite{Al-Badawi2024CoTPh..76h5401A,Al-Badawi12024,Al-Badawi2024b,Alloqulov12024,Alloqulov2025}
\begin{equation}
\hat{\alpha}_{SIS}=\hat{\alpha}_{SIS1}+\hat{\alpha}_{SIS2}+\hat{\alpha}_{SIS3} \label{nonsis}\ .
\end{equation}

From Eqs.~(\ref{h}), (\ref{anglemain}), and (\ref{nonsis}), we can analytically derive the deflection angle for BH surrounded by SIS as

\begin{figure*}[!htb]
 \centering
    \includegraphics[scale=0.5]{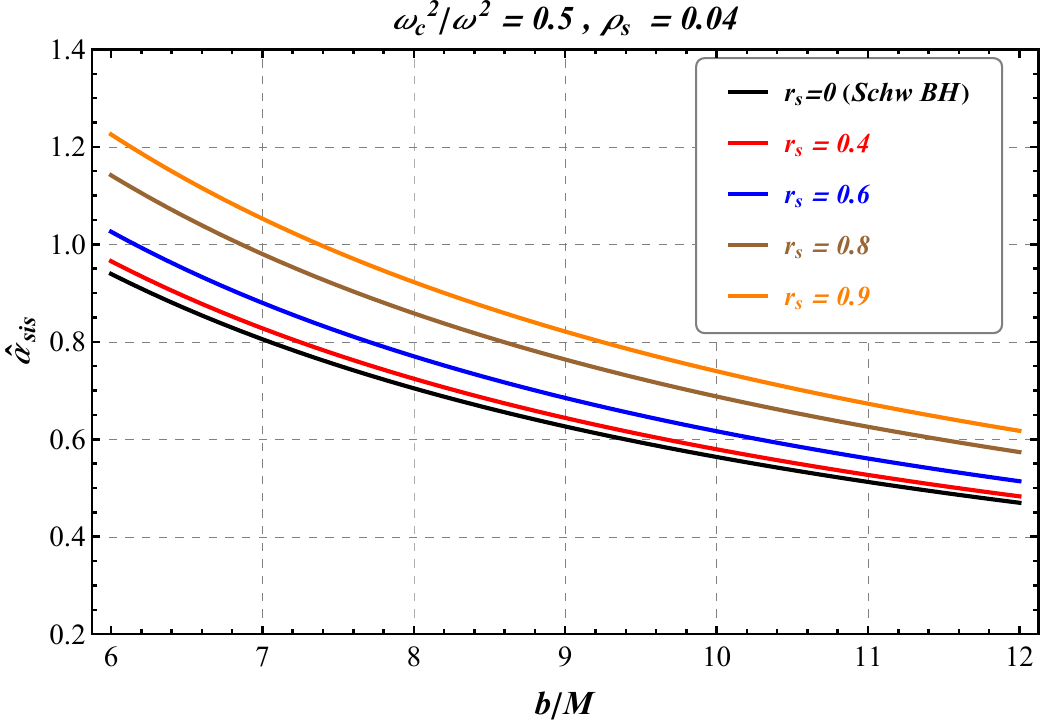}
     \includegraphics[scale=0.5]{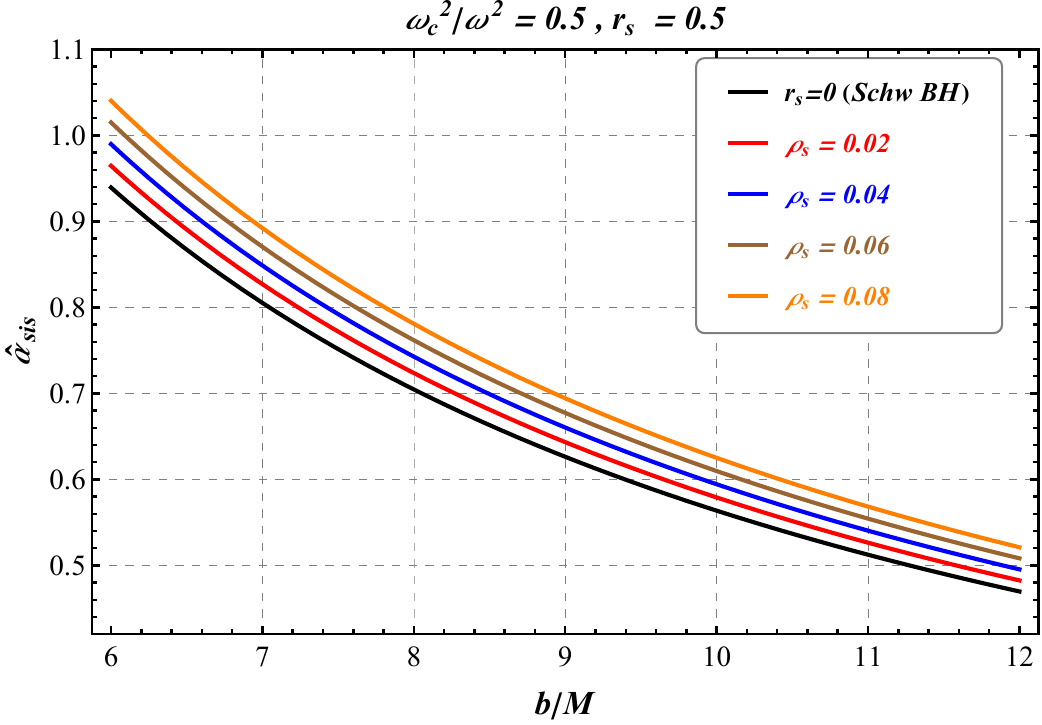}
      \caption{The variation of $\hat{\alpha }_{\text{sis}}$ as a function of the parameter $b/M$ for different values of scale radius (left panel) and density of DM halo (right panel). }s\label{a2}
     \end{figure*}

\begin{widetext}
\begin{align}
   \hat{\alpha}_{SIS}=&\frac{4 M}{b}+4 \pi ^2 r_s^2 \rho _s+\frac{8 \pi  r_s^3 \rho _s}{b}-\frac{32 M^2 \omega _c^2 r_s \rho _s}{b \omega ^2}+\frac{8 \pi  M^2 \omega _c^2 r_s^2 \rho _s}{b^2 \omega ^2}+\frac{2 M^2 \omega _c^2}{\pi  b^2 \omega ^2}+\frac{16 M^3 \omega _c^2}{3 \pi  b^3 \omega ^2}\nonumber\\&+\frac{16 \pi  M^2 \omega _c^2 \rho _s}{\omega ^2}\Bigg[1+\frac{ b \left(\tan ^{-1}\left(\frac{r_s}{\sqrt{b^2-r_s^2}}\right)-\tan ^{-1}\left(\frac{b+r_s}{\sqrt{b^2-r_s^2}}\right)\right) \left(\frac{\omega ^2 r_s^2}{M^2 \omega _c^2}+\frac{4}{\pi }\right)}{\sqrt{b^2-r_s^2}}\Bigg]\ ,
  \label{eq:a_sis}
  \end{align}
   \end{widetext}
One can write the explicit form of the plasma frequency as~\cite{Bisnovatyi-Kogan2010}
\begin{equation}
\omega^2_c=\frac{K_e \sigma^2_{\nu}}{2\pi \kappa m_p R^2_S}\ , 
\end{equation}
From Eq.\eqref{eq:a_sis}, we analyze how the deflection angle $\hat{\alpha}_{SIS}$ varies with the parameter $b$ for different combinations of $r_s$, $\rho_s$, ${\omega^2_{c}/\omega^2}$ in a non-uniform plasma medium (see Fig. \ref{a2}). The deflection angle shows a similar trend to that found in the uniform plasma case, as illustrated in Fig.~\ref{a2}. The deflection angle demonstrates a clear difference in magnitude: it is larger in a uniform plasma medium compared to a non\-uniform plasma medium, as seen in Figs.~\ref{a1} and \ref{a2}.

\section{{Magnification of gravitationally lensed image}
}\label{Sec:magnification}

In this section, we present an analysis of the magnification for the gravitationally lensed image of a BH surrounded by a DM halo using the derived deflection angle. When the effect of the DM halo is taken into account, the lens equation can be expressed in terms of the relevant angles, namely $\hat{\alpha}$, $\theta$, and $\beta$ (see e.g.,~\cite{Bozza2008})
\begin{align}\label{lenseq}
\theta D_\mathrm{s}=\beta D_\mathrm{s}+\hat{\alpha}D_\mathrm{ds}\ . 
\end{align}
Here, \( D_s \) is the distance from the source to the observer, while \( D_{ds} \) corresponds to the distance between the source and the lens. $\beta$ denotes the angular position of the source, and $\theta$  is the image angle.

A circular image produced by lensing is called an Einstein ring, and the corresponding angular radius $\theta_0$ of this ring~\cite{Schneider1992} is given by 
\begin{align}
\theta_0=\sqrt{\frac{2R_s D_{ds}}{D_dD_s}}\ .
\label{Einstein}\end{align}
where $R_s=2M$, and $D_d$ is the distance of the lens-observer.

The magnification is calculated from the ratio of the total lensed intensity $I_{tot}$ to the original unlensed intensity $I_*$~\cite{Schneider1992}
\begin{align}\label{magni}
\mu_{\Sigma}=\frac{I_\mathrm{tot}}{I_*}=\underset{k}\sum\bigg|\bigg(\frac{\theta_k}{\beta}\bigg)\bigg(\frac{d\theta_k}{d\beta}\bigg)\bigg|, \quad k=1,2, \dotsc,  j\ ,
\end{align}
 The magnifications of the source, corresponding to the outer and inner images, and their total magnification $\mu_{tot}^{pl}$ can be expressed as follows~\cite{Bozza2008,Schneider1992,Bisnovatyi-Kogan2010}
\begin{align}\label{postparity}
\mu^\mathrm{pl}_\mathrm{+}=\frac{1}{4}\bigg(\frac{x}{\sqrt{x^2+4}}+\frac{\sqrt{x^2+4}}{x}+2\bigg)\ ,
\end{align}
\begin{align}\label{negparity}
\mu^\mathrm{pl}_\mathrm{-}=\frac{1}{4}\bigg(\frac{x}{\sqrt{x^2+4}}+\frac{\sqrt{x^2+4}}{x}-2\bigg)\ ,
\end{align}
\begin{align}\label{magtot}
\mu^\mathrm{pl}_\mathrm{tot}=\mu^\mathrm{pl}_{+}+\mu^\mathrm{pl}_{-}=\frac{x^2+2}{x\sqrt{x^2+4}}\ .
\end{align}
Here, the ratio $x={\beta}/{\theta_0}$ represents a dimensionless parameter.

We now analyze the magnification behavior around a BH with the DM halo, considering two types of plasma distribution.

\textit{Uniform plasma case}: As discussed in Section~\ref{sec:plasma} , using the deflection angle $\alpha_{uni}$ obtained there, we analyze the influence of uniform plasma on the magnification around a BH with the DM halo.
Consequently, Eq.~\eqref{lenseq} can be written in the following form (see, e.g.~\cite{Bisnovatyi-Kogan2010})
\begin{align}\label{newlenseq2}
\beta=\theta -\frac{D_\mathrm{ds}}{D_\mathrm{s}}\hat{\alpha}_{uni}=\theta-(\theta_0^{pl})_{uni}\ ,
\end{align}
where
\begin{align}
(\theta_0^{pl})_{uni}=\frac{D_\mathrm{ds}}{D_\mathrm{s}}\hat{\alpha}_{uni}\ .
\label{theta}
\end{align}
Let us assume that the deflection angle is given by the following expression:
\begin{align}
\hat{\alpha}_{uni}=\frac{4M}{b}(1+A_{uni})=\frac{2R_s}{D_d(\theta_0^{pl})_{uni}}(1+A_{uni})\ .
\end{align}
Here, $R_s=2M$, $b=D_\mathrm{d}\theta^{pl}_0$, and  $A_{uni}$ represents the additional contribution to the deflection angle arising from both plasma and DM halo effects, and this quantity is computed numerically. Then, substituting the above equation and Eq.~\eqref{Einstein} into Eq.~\eqref{theta} yields the following:
\begin{align}
(\theta_0^{pl})_{uni}&=&&\frac{D_\mathrm{ds}}{D_\mathrm{s}}\hat{\alpha}_{uni}=\frac{2R_s D_\mathrm{ds}}{D_dD_\mathrm{s}}\frac{1}{(\theta_0^{pl})_{uni}}(1+A_{uni})\nonumber\\&=&&\frac{\theta_0^2}{(\theta_0^{pl})_{uni}}(1+A_{uni})\ ,
\end{align}
\begin{align}
\frac{(\theta_0^{pl})_{uni}}{\theta_0}=\sqrt{1+A_{uni}}\ ,
\end{align}
from this, we find the ratio of dimensionless quantities:
\begin{eqnarray}
    x_{uni}/x_0=\frac{\beta}{(\theta_0^{pl})_{uni}}/\frac{\beta}{\theta_0}=\frac{\theta_0}{(\theta_0^{pl})_{uni}}=\frac{1}{\sqrt{1+A_{uni}}}\ ,
\end{eqnarray}
\begin{eqnarray}
    x_{uni}=\frac{x_0}{\sqrt{1+A_{uni}}}\ .
\end{eqnarray}
Here, $x_0$ represents the normalized angle in the absence of both plasma and the DM halo.

We then rewrite the magnifications given by Eqs.~\eqref{postparity},~\eqref{negparity}, and ~\eqref{magtot} for the case of a uniform plasma
\begin{figure*}[!htb]
 \centering
    \includegraphics[scale=0.4]{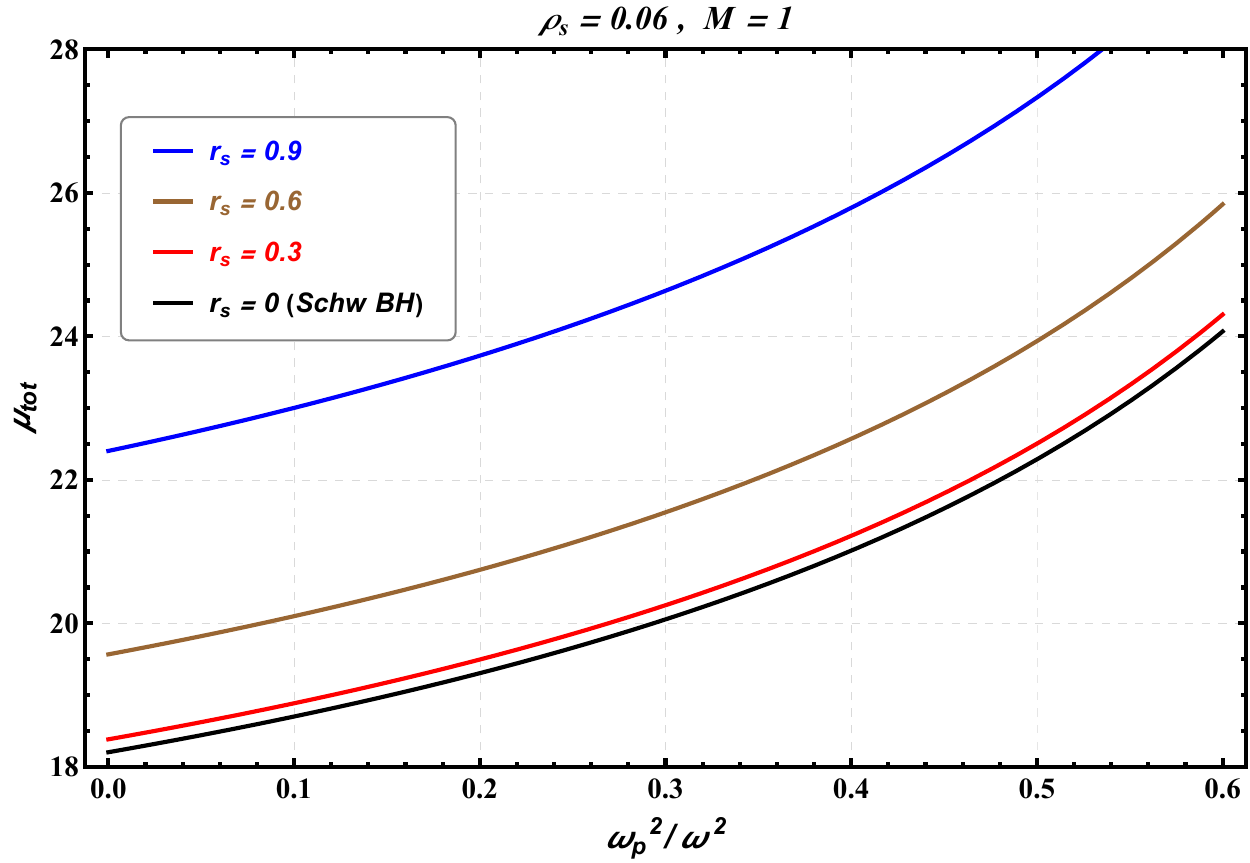}
     \includegraphics[scale=0.4]{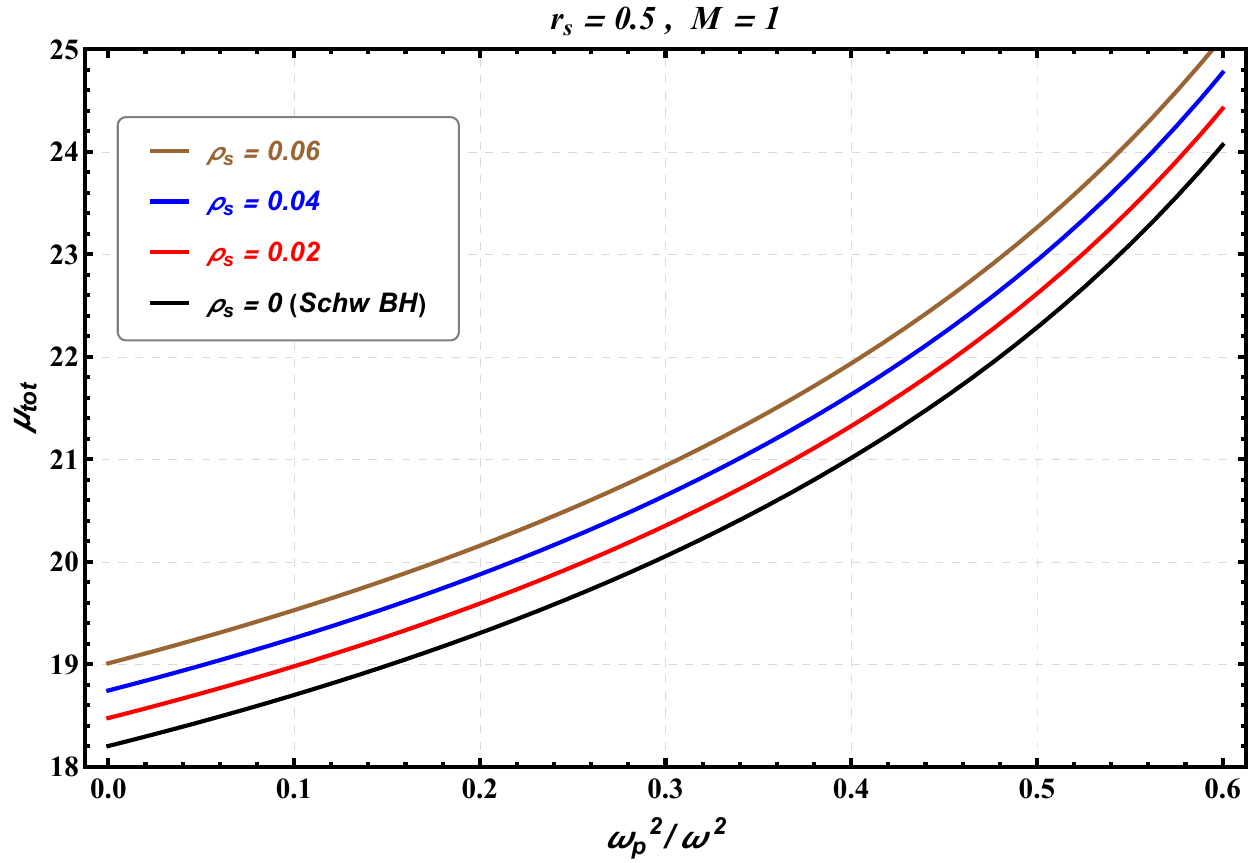}
      \caption{Plot shows the dependence of $\mu_{\text{tot}}$ on $\omega_p^{2}/\omega^{2}$ for different values of the scale radius $r_s$ (left panel) and characteristic density $\rho_s$ (right panel). In both plots, we fix $b/M=5$.}\label{fig:mu}
     \end{figure*}
\begin{equation}
       (\mu^{pl}_+)_{uni}=\frac{1}{4}\left(\dfrac{x_{uni}}{\sqrt{x^2_{uni}+4}}+\dfrac{\sqrt{x^2_{uni}+4}}{x_{uni}}+2\right)\ ,
\end{equation}
\begin{equation}
       (\mu^{pl}_-)_{uni}=\frac{1}{4}\left(\dfrac{x_{uni}}{\sqrt{x^2_{uni}+4}}+\dfrac{\sqrt{x^2_{uni}+4}}{x_{uni}}-2\right)\ ,
\end{equation}
\begin{equation}
(\mu^{pl}_{tot})_{uni}=\dfrac{x^2_{uni}+2}{x_{uni}\sqrt{x^2_{uni}+4}}\ .
\end{equation}

From the above relation, the total magnification is determined as a function of the normalized plasma frequency $\omega^2_p/\omega^2$ for a BH embedded in a DM halo within a uniform plasma medium, as shown in Fig.~\ref{fig:mu}. Here, the normalized angle $x_0$ is fixed. From the left panel, it is clearly seen that as the characteristic scale $r_s$ increases, the magnification curve shifts upward relative to the Schwarzschild BH case ($r_s=0$, $\rho_s=0$). When the characteristic scale $r_s$ is increased in equal increments, the increase in magnification at each step becomes roughly more than twice that of the preceding step. In the right panel, the scale radius $r_s$ is fixed, and as the characteristic density $\rho_s$ is increased in equal steps, the magnification curve shifts upward by nearly the same amount at each increment.
\begin{figure*}[!htb]
 \begin{center}
   \includegraphics[scale=0.4]{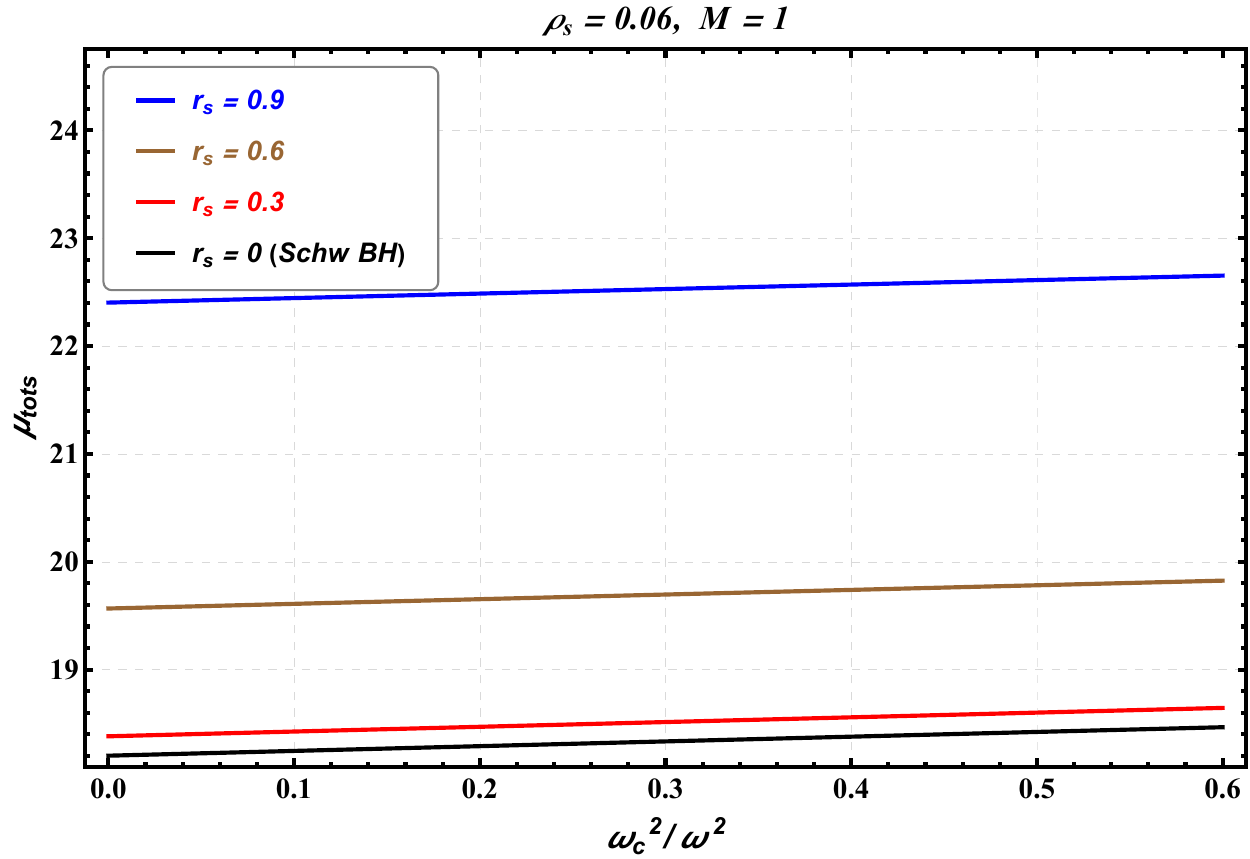}
   \includegraphics[scale=0.4]{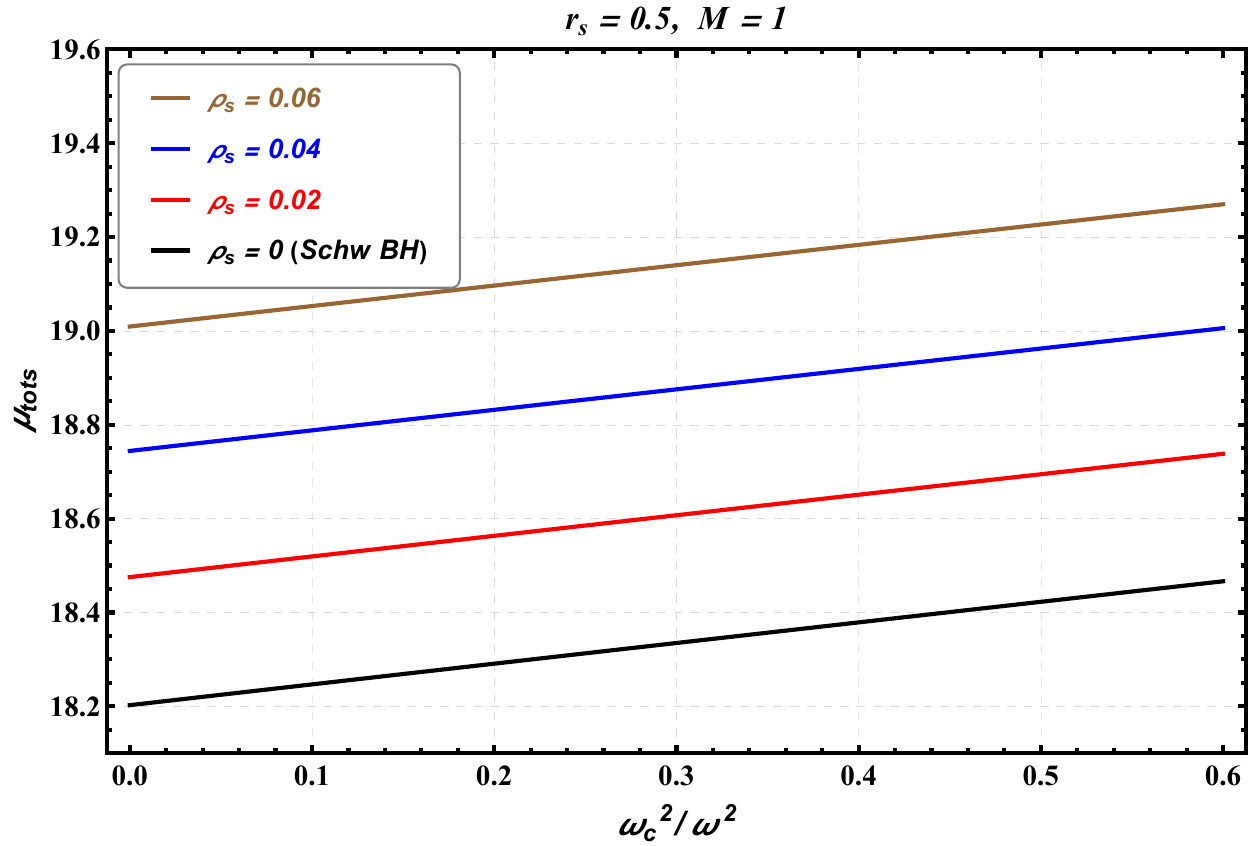}
  \end{center}
\caption{The plots illustrate the behaviour of $\mu_{tots}$ as a function of the normalized SIS plasma frequency $\omega^2_c/\omega^2$ for different values of DM halo parameter, with $b/M=5$. Here, in the left panel, the characteristic density is fixed at $\rho_s = 0.04$, and in the right panel, the scale radius is set to $r_s = 0.5$.}\label{fig:ms}
\end{figure*}

\textit{Non-uniform plasma case}: For the SIS model, we perform calculations similar to those above. To do so, we use the deflection angle $\hat{\alpha}_{SIS}$ given in Eq.~\eqref{eq:a_sis} and derive the expression for the normalized angle:
\begin{widetext}
\begin{align}
  x_{SIS}=\frac{\beta}{(\theta^{pl}_0)_{SIS}}=x_0\frac{\theta_0}{(\theta^{pl}_0)_{SIS}}=x_0\Bigg[1+\frac{2 \pi  r_s^3 \rho _s}{M}+\frac{\pi ^2 b r_s^2 \rho _s}{M}-\frac{8 M \omega _c^2 r_s \rho _s}{\omega ^2}+\frac{4 \pi  b M \omega _c^2 \rho _s}{\omega ^2}+\frac{2 \pi  M \omega _c^2 r_s^2 \rho _s}{b \omega ^2}+\frac{M \omega _c^2}{2 \pi  b \omega ^2}\nonumber\\+\frac{4 M^2 \omega _c^2}{3 \pi  b^2 \omega ^2}+\frac{4 b^2 \rho _s \left(\tan ^{-1}\left(\frac{r_s}{\sqrt{b^2-r_s^2}}\right)-\tan ^{-1}\left(\frac{b+r_s}{\sqrt{b^2-r_s^2}}\right)\right) \left(\frac{4 M^2 \omega _c^2}{\omega ^2}+\pi  r_s^2\right)}{M \sqrt{b^2-r_s^2}}\Bigg]^{-1/2}\ ,
\end{align}
\end{widetext}

Finally, by substituting the normalized angle obtained for the SIS model into Eq.~\eqref{magtot}, we determine the total magnification:
\begin{equation}
(\mu^{pl}_{tot})_{SIS}=\dfrac{x^2_{SIS}+2}{x_{SIS}\sqrt{x^2_{SIS}+4}}\ .\label{eq:magsis}
\end{equation}

Fig.~\ref{fig:ms} displays the relationship between $\mu_{tot}$ and the normalized SIS plasma frequency $\omega^2_c/\omega^2$ for a BH with the DM halo. In the left panel, the characteristic density is fixed at $\rho=0.06$, and each colored curve represents a different scale {radius}, $r_s=0,0.3,0.6,0.9$, with the black curve corresponding to the Schwarzschild BH case (Schw BH). It is evident that as the scale radius $r_s$ increases, the {parameter} $\mu_{tots}$ also increases. In the right panel, the scale radius is held fixed at $r_s=0.5$, while the characteristic density $\rho_s$ is varied. The magnification curves shift upward with increasing $\rho_s$, indicating that halos with higher density produce greater magnification. 

\section{Conclusion}
\label{Sec:conclusion}

In this paper, we probed the optical phenomena observed near a Schwarzschild-like BH with the Dehnen-type (1,4,2) DM halo. First, we began to examine the spacetime metric and the null geodesic paths. We showed that the radius of the photon sphere increases as both the characteristic density $\rho_s$ and the scale radius $r_s$ of the DM halo grow, resulting in a stronger
gravitational effect (see Fig.~\ref{fig2}). Further, in the weak-field regime, we applied the Gauss–Bonnet (GB) theorem and obtained an analytical expression for the deflection angle. The analysis reveals that increasing the characteristic density $\rho_s$ and the scale radius $r_s$ of the DM halo increases the magnitude of the deflection angle, as shown in Fig.~\ref{fig:delta}.

Additionally, we explored lensing effects by solving the deflection-angle integral numerically to obtain the photon trajectories and the corresponding ray-tracing plots near a Schwarzschild-like BH with the DM halo. We showed that the gravitational field of the BH-DM system strengthens as both DM halo parameters increase, causing the deflection angle to increase, as clearly illustrated in Figs.~\ref{fig:ray} and~\ref{fig:ray2}. This happens because the critical impact parameter shifted from $b_c = 5.196$ for a Schwarzschild BH to $b_c = 5.584$ for a BH with the DM halo (i.e., for $r_s=0.6$ and $\rho_s=0.03$). The lensing ring intervals shifted towards larger values, indicating that both the characteristic density and the scale parameters enhance the gravitational field (see Table~\ref{tab:nb}).

We also analyzed photon trajectories for both uniform and non-uniform plasma distributions. To this end, we derived the corresponding equations of motion and determined the photon sphere radius as well as the resulting BH shadow radius. As expected, in addition to the effects of the parameters $r_s$ and $\rho_s$, the presence of a plasma medium has a significant impact on the the photon sphere, i.e., lower plasma densities lead to a larger photon-sphere radius. Similarly, the radius of the BH shadow increases as $r_s$ and $\rho_s$ grow (see Fig. \ref{fig.7}). Observational bounds on the DM halo parameters $r_s$ and $\rho_s$ within a plasma environment were obtained using observational data of M87* and Sgr A* (see Figs.~\ref{fig.9} and \ref{fig.8}). Finally, we found that both the DM halo scale parameter $r_s$ and the halo density $\rho_s$ increase the deflection angle $\hat{\alpha }$ (see Figs.~\ref{a1} and \ref{a2}).

The plasma medium significantly influences gravitational lensing: the deflection angle is larger for a uniform plasma case compared to a SIS plasma case, highlighting the strong dependence of lensing effects on the specific plasma distribution. In the uniform plasma case, we showed that increasing the scale radius $r_s$ and characteristic density $\rho_s$ produces an overall enhancement of magnification. Similarly, in the SIS plasma, larger $r_s$ or $\rho_s$ produce higher magnification compared to the Schwarzschild BH case, indicating that denser or more extended DM halos amplify the observed lensing features (see Figs.~\ref{fig:mu} and~\ref{fig:ms}).

Our findings show that the Schwarzschild-like BH with the Dehnen-type DM halo produced larger photon sphere radii, stronger deflection angles in both weak and strong lensing, larger BH shadow radii, and noticeable modifications of lensing and magnification in both uniform and SIS plasma environments than in the Schwarzschild BH case. The surrounding DM halo and the presence of the plasma make this study astrophysically relevant and important for understanding the optical properties observed near BHs.

\begin{acknowledgements}
PS acknowledge the support of the Anusandhan National Research Foundation (ANRF) under the Science and Engineering Research Board (SERB) Core Research Grant (Grant No.\ CRG/2023/008980). 
\end{acknowledgements}

\bibliography{references,ref2, ref5}
\end{document}